\documentclass[structabstract]{aa}  
\usepackage{amssymb}
\usepackage{amsmath}
\usepackage{amsfonts}
\usepackage[latin1]{inputenc}
\usepackage{graphicx}
\usepackage{epsfig}
\usepackage{subfigure}
\usepackage[authoryear]{natbib}
\usepackage[figuresright]{rotating}
\usepackage{supertabular}
\usepackage{longtable}
\usepackage{lscape}
\usepackage{txfonts}
\newcommand{\teff}{\ifmmode T_{\rm eff} \else $T_{\mathrm{eff}}$\fi}
\newcommand{\logg}{\ifmmode \log g \else $\log g$\fi}
\newcommand{\lL}{\ifmmode \log \frac{L}{L_{\odot}} \else $\log \frac{L}{L_{\odot}}$\fi}

\newcommand{\vsini}{\ifmmode v \sin i \else $v \sin i$\fi}

\newcommand{\kms}{km~s$^{-1}$}
\newcommand{\msun}{\ifmmode M_{\odot} \else $M_{\odot}$\fi}
\newcommand{\zsun}{\ifmmode Z_{\odot} \else $Z_{\odot}$\fi}
\newcommand{\lsun}{\ifmmode L_{\odot} \else $L_{\odot}$\fi}
\newcommand{\rsun}{\ifmmode R_{\odot} \else $R_{\odot}$\fi}
\newcommand{\qh}{\ifmmode Q_{\rm H} \else $Q_{\rm H}$\fi}
\newcommand{\qhei}{\ifmmode Q_{\ion{He}{i}} \else $Q_{\ion{He}{i}}$\fi}

\begin{document}
   \title{A spectroscopic investigation of the O-type star population in four Cygnus OB associations}

   \subtitle{ I. Determination of the binary fraction}

   \author{L. Mahy\inst{1}
          \and
          G. Rauw\inst{1}
          \and
          M. De Becker\inst{1}
          \and
          P. Eenens\inst{2}
          \and
          C. A. Flores\inst{2}
          }

   \offprints{L. Mahy}

   \institute{
     Institut d'Astrophysique et de G\'eophysique, Universit\'e de Li\`ege, B\^at. B5C, All\'ee du 6 Ao\^ut 17, B-4000, Li\`ege, Belgium\\
     \email{mahy@astro.ulg.ac.be}
     \and 
     Departamento de Astronom\'ia, Universidad de Guanajuato, Apartado 144, 36000 Guanajuato, GTO, Mexico
   }
   
   \date{Received ...; accepted ...}
   
 
  \abstract
      {Establishing the multiplicity of O-type stars is the first step towards accurately determining their stellar parameters. Moreover, the distribution of the orbital parameters provides observational clues to the way that O-type stars form and to the interactions during their evolution.}
   {Our objective is to constrain the multiplicity of a sample of O-type stars belonging to poorly investigated OB associations in the Cygnus complex and for the first time to provide orbital parameters for binaries identified in our sample. Such information is relevant to addressing the issue of the binarity in the context of O-type star formation scenarios.}
   {We performed a long-term spectroscopic survey of nineteen O-type stars. We searched for radial velocity variations to unveil binaries on timescales from a few days up to a few years, on the basis of a large set of optical spectra.}
   {We confirm the binarity for four objects: HD\,193443, HD\,228989, HD\,229234 and HD\,194649. We derive for the first time the orbital solutions of three systems, and we confirm the values of the fourth, showing that these four systems all have orbital periods shorter than 10 days. Besides these results, we also detect several objects that show non-periodic line profile variations in some of their spectral lines. These variations mainly occur in the spectral lines, that are generally affected by the stellar wind and are not likely to be related to binarity.}
   {The minimal binary fraction in our sample is estimated to be 21\%, but it varies from one OB association to the next. Indeed, 3 O stars of our sample out of 9 (33\%) belonging to Cyg\,OB1 are binary systems, 0\% (0 out of 4) in Cyg\,OB3, 0\% (0 out of 3) in Cyg\,OB8, and 33\% (1 out of 3) in Cyg\,OB9. Our spectroscopic investigation also stresses the absence of long-period systems among the stars in our sample. This result contrasts with the case of the O-type stellar population in NGC\,2244 among which no object showed radial velocity variations on short timescales. However, we show that it is probably an effect of the sample and that this difference does not a priori suggest a somewhat different star forming process in these two environments.}

   \titlerunning{The binary fraction of O-type star population in four Cygnus OB associations}
   \authorrunning{L. Mahy et al.}
   \maketitle


\section{Introduction}
\label{sect:intro}
The multiplicity of massive stars constitutes one of the most essential ingredients for understanding these objects. Establishing it is the first step in obtaining valuable physical parameters, such as their masses, radii and luminosities. The binarity also affects the evolutionary paths of the components in such systems in comparison to single stars through phenomena, such as tidal interations or Roche lobe overflows \citep{sanaevans2010}. Knowing this multiplicity also allows us to be more accurate on the wind parameters, especially the mass-loss rates, or on the intrinsic X-ray luminosities. Moreover, the distribution of the orbital parameters for the massive systems provides us with additional information on the way that massive close binaries are formed. 

Many different scenarios for the formation of these objects have been proposed \citep[see][for a complete review]{zy07}, but none of them totally explains what happens during their formation or during its earliest stages. This renders our knowledge of massive stars still fragmentary. In addition, the distribution of orbital parameters allows us to give insight into other avertions such as whether the binary fraction of massive stars can be related to the stellar density \citep{pen93,gar01}. The orbital parameters of multiple systems can thus provide important clues to the conditions prevailing in the birthplace during their formation or to the dynamical interations that occur during the earliest stages of their evolution. Several studies have already been performed in spectroscopy to determine the binary fraction of O-type stars in nearby young open clusters \citep[see e.g.,][and references therein]{sanaevans2010}. All these analyses revealed an averaged binary fraction of about $44\pm5$\% \citep{sanaevans2010}. However, to investigate such a large parameter space, it is necessary to combine other observation techniques, such as interferometry \citep{nelan2004}, speckle interferometry \citep[see e.g., ][]{mas98, mai04, mas09}, or adaptative optics \citep{tur08}. Indeed, all these large surveys give the advantages of better statistics whilst the surveys devoted to young open clusters or OB associations allow more homogeneous stellar populations to be investigated as noted by \citet{kim12}.

\begin{figure}[htbp]
\begin{center}
\includegraphics[width=8.cm,bb=10 3 535 355,clip]{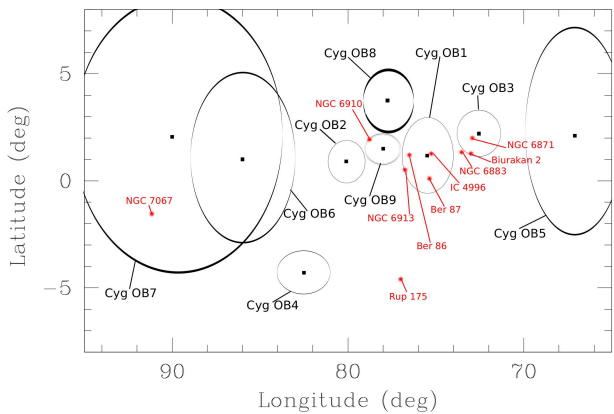}
\caption{Schematic view of the different OB associations and young open clusters in the Cygnus region.}\label{fig:plan}
\end{center}
\end{figure}

The Cygnus area is an active star-forming region in the Milky Way. Its relative proximity ($0.7-2.5$~kpc, \citealt{uyaniker2001}) makes this region suitable for studying stellar populations and star formation processes \citep[see][for a review]{cygnusstarform}. The Cygnus constellation harbours nine OB associations and at least a dozen of young open clusters. To better constrain the environment of the Cygnus region, Fig.\ref{fig:plan} represents a schematic view with the locations, in galactic coordinates, of the different OB associations and of several young open clusters. The central association Cyg\,OB2 is certainly the most famous and one of the youngest of the Cygnus region with Cyg\,OB8. To be more accurate, Cyg\,OB2 has in fact been shown to possibly have two populations with ages of $2-3.5$ and $5$~Myrs \citep{wri10}, whilst the O stars seem to belong to a younger population, possibly aged about $2$~Myrs \citep{Han03}. In its surroundings, the associations Cyg\,OB1, Cyg\,OB3, Cyg\,OB7 and Cyg\,OB9 are older with ages of about 7.5, 8.3, 13.0 and 8.0 Myrs, respectively \citep{uyaniker2001}. Though the Cygnus region is rich in O-type stars, its heavy absorption prevents one from obtaining a better estimate of the stellar population even though about 100 hot stars are expected in Cyg\,OB1, Cyg\,OB8 and Cyg\,OB9 together \citep{mel95} and about 100 O-stars are expected only in the Cyg\,OB2 association \citep{kno2002,com2002}. However, the membership of certain massive stars to a given cluster or association is sometimes not clearly defined due to the complexity of this area. Recent investigation of the multiplicity of massive stars in Cyg\,OB2 association performed by \citet[][and subsequent papers]{kim12b} has shown that the hard minimum binary fraction for the massive stars in this association is estimated at 21\%. Such an investigation for the massive star population in the other OB associations in the Cygnus complex is currently lacking.

We have undertaken a long-term spectroscopic survey of nineteen O-type stars sampled over four different OB associations (Cyg\,OB1, Cyg\,OB3, Cyg\,OB8 and Cyg\,OB9) as well as in several young open clusters belonging to these associations (e.g., Berkeley\,86, NGC\,6871, NGC\,6913,...). Their membership to a given association or cluster is established on the basis of the catalogue of \citet{hum78}. However, many other O-type stars belong to these associations but we only focus on these targets because they are relatively bright and their binary status was not yet subject to a detailed spectroscopic study, except for HD\,229234. We will also include other stars already known as binaries in the discussions on the binary fraction in the different Cygnus OB associations. Moreover, the majority of the objects investigated in the present paper were poorly observed in the past mainly due to the high extinction towards the Cygnus region. In this context, this paper (the first of a series of two) aims at establishing the multiplicity as well as the spectral classification of these stars. This analysis focuses on other regions in the Cygnus complex than the large survey of \citet[][and the subsequent papers]{kim12b} which was devoted to the Cyg\,OB2 association. Therefore, this analysis can thus be seen as a complement to the study of the binary fraction in Cygnus region.

The present paper is organized as follows. The observing campaign is described in Sect.~\ref{sect:obs}. The binary status of the stars in our sample is discussed in the next four sections according to the different OB associations. Section~\ref{sect:ob1} is devoted to O-type stars in the Cyg\,OB1 association, Sect.~\ref{sect:ob3} to these objects in Cyg\,OB3, Sect.~\ref{sect:ob8} to Cyg\,OB8, and Sect.~\ref{sect:ob9} to Cyg\,OB9. Finally, Sect.~\ref{sect:disc} discusses the observational biases and the binary fraction that we determine whilst Sect.~\ref{sect:conc} summarizes our results.  
 

\section{Observations and binary criteria}
\label{sect:obs}

Deriving the binary fraction of a given population requires an intense survey of the stars but also a good time sampling to constrain the short- as well as the long-period systems. This monitoring has to be as homogeneous as possible because combining data from the literature may provide erroneous results. Indeed, the literature (especially the older papers) often does not specify on which spectral line the radial velocities (RVs) were measured. Therefore, large differences can be observed between RVs measured on two different lines of a presumably single star. This can lead to some overdeterminations of the spectroscopic binary fraction as notably observed in \citet{gar01} for the young open cluster IC\,1805 \citep{rau04,deb06,hillwig2006}. Moreover, it is important to keep in mind the possible observational biases that can prevent the detection of a binary system. Therefore, stars for which no evidence of binarity is found can never be definitely considered as single. Even if no RV shift is detected, the system could be seen under a particular orientation, have a very long period or perhaps a high eccentricity, thereby making the RV variations not significant over a long timescale.

We performed a spectroscopic survey spread over three years (from 2008 to 2011) of all the nineteen stars of our sample, except for HD\,194280 which was observed between 2004 and 2007 (see Table\,\ref{list}). This selection of stars however constitutes another observational bias because we limited our observations to stars bright enough to yield good quality data with exposure times not longer than one hour with 1.5 or 2-\,meter class telescopes. Considering the high -- and heterogeneous -- extinction towards the Cygnus region, the present monitoring of O-type stars (among which some could also be member of multiple systems) is far from being complete. Nevertheless, we collected a set of 274 spectra for these nineteen stars by using three different instruments. Table~\ref{list} lists the stars and the parameters of the observing runs. Table~\ref{rv}, available electronically, gives the Heliocentric Julian Dates (HJD) and the RVs measured on different spectral lines for all the observations. For the SB2 systems, the RVs refined by cross-correlation are also provided. In the latter table as in the following of the present paper, HJD is expressed in format $\mathrm{HJD}-2\,450\,000$ for readability. Our dataset thus allowed us to study the RV and line profile variations in the observed spectra over short- (a few days) and long-term (a few years) timescales.

A first part of our dataset was obtained with the Aur{\'e}lie spectrograph at the 1.52m telescope at Observatoire de Haute-Provence (OHP, France). The spectra are obtained with a resolving power of about $R = 9\,000$, they are centred on 4650~\AA\ and cover wavelengths between 4450~\AA\ and 4900~\AA. However, for some spectra of HD\,193443, another configuration of the instrument was chosen. In this case, the resolving power was equal to $R=12000$, these spectra are centred on 4550~\AA\ and cover 4450 to 4650~\AA. Typical exposure times between 25 and 60 min yielded spectra with a signal-to-noise ratio larger than 150 as measured in the continuum close to 4800~\AA. The data reduction procedure, performed with the \rm{MIDAS} software, is described in \citet{rau04}.

Another part of the data was collected with the ESPRESSO spectrograph mounted on the 2.12m telescope at Observatorio Astron{\'o}mico Nacional of San Pedro M{\`a}rtir (SPM) in Mexico. This échelle spectrograph provides spectra spread over 27 orders, covering a wavelength domain between 3780~\AA\ and 6950~\AA\ with a spectral resolving power of $R = 18\,000$. Typical exposure times ranged from 15 to 30 min. We combined the consecutive exposures of an object in a given night to obtain signal-to-noise ratios close to 150. As a result, our time series did not allow us to investigate RV variations with timescales shorter than $0.8$ day, which is not a problem for multiplicity investigations as typically the shortest periods for massive stars are longer than one day (as Kepler's third law would imply orbital separations smaller than the radii of the components for such short periods). The data were reduced using the échelle package available within the \rm{MIDAS} software.

 We also retrieved eight spectra taken before 2008 from the Elodie archives to complete our dataset. Elodie was an {\'e}chelle spectrograph, mounted at the 1.93m telescope at OHP, with 67 orders in the [$3850-6850$]~\AA\ spectral domain and a resolving power of $R = 42\,000$.

We fit one or two Gaussian profiles on several strong spectral lines to estimate the RVs of the stars when the line profiles were not too much broadened by rotation. For the rapid rotators ($\vsini~\geq 300$~\kms), we used a synthetic line profile broadened with the projected rotational velocity of the star to measure the Doppler shifts by least square. This RV analysis is performed by using the rest wavelengths quoted by \citet{con77} for the [$4000-5000$]~\AA\ wavelength domain and \citet{und94} for wavelengths above 5000~\AA. The common wavelength region between the majority of the data is [$4450-4900$]~\AA. In this spectral band, some interstellar lines exist but they are not sufficiently strong and narrow to use them as indicators of the RV uncertainties. Our previous experience with the same spectrographs \citep[see][]{mah09} showed that a standard deviation on the RVs of about $7-8$~\kms\ is commonly achieved with these spectra. We therefore consider the RV changes as significant when the standard deviations computed from the RV measurements reach at least this value. For rapid rotators, we consider them as variable if standard deviations of at least 15~\kms\ are determined. We stress that the threshold value for the rapid rotators has been increased relative to \citet{mah09} because the present data are slightly noisier. As a consequence, we consider, in the present paper, that a star is an SB1 binary system if the RV variations are significant (i.e., the standard deviation is at least larger than $7-8$~\kms\ for the slow and mid rotators and at least 15~\kms\ for rapid rotators) and periodic. We are dealing with an SB2 if the signature of a companion whose spectral lines move in anti-phase with those of the main star is observed. Moreover, if in the cases where significant variations of the RVs are measured on all the spectral lines without any periodic motion, the object will be labelled as a ``binary candidate''. Finally, we also calculate the Temporal Variance Spectrum \citep[TVS,][]{ful96} to search for line profile variations. The TVS provides quantitative information of the temporal variability as a function of the wavelength. If the variations exceed the threshold we can reject the null hypothesis of non-variability of the spectral lines. When such variations are detected for {\it all} the line profiles, the TVS may then be considered as an additional indicator of multiplicity and, in this case, the star is reported as binary candidate. 

The SB2 binary systems detected in the present work are analysed as follows. After having measured the RVs of both components, we use these values as an input of our disentangling programme. This programme is based on the \citet{gl06} technique and it allows us to compute the individual spectra of both components, and to refine the RVs by applying a cross-correlation technique. It is thus possible to obtain more reliable RVs at phases where the spectral lines are heavily blended. The cross-correlation windows used for the RV determinations generally gather the strongest helium lines (\ion{He}{i}~4471, \ion{He}{ii}~4542, \ion{He}{ii}~4686 and \ion{He}{i}~4713) for both components. With these refined RVs, we then apply the Heck-Manfroid-Mersch method (hereafter HMM, \citealt{hec85}, revised by \citealt{gos01}) to the time series of $RV_{\mathrm{S}}-RV_{\mathrm{P}}$ to determine the orbital period of the system. The 1-$\sigma$ error-bar on the period is estimated by assuming an uncertainty on the frequency of the peak associated to the orbital period equal to 10\,$\%$ of the natural width of the peaks in the Fourier power spectrum (e.g., $\Delta\nu \sim (10\,\Delta T_{\mathrm{tot}}$)$^{-1}$, where $\Delta T_{\mathrm{tot}}$ is the time elapsed between the first and the last observations of our campaign). The orbital period is then used to determine the orbital solution of the system by applying the Li{\`e}ge Orbital Solution Package (LOSP\footnote{This program, maintained by H. Sana, is available at http://www.science.uva.nl/$\sim$hsana/losp.html and is based on the generalization of the SB1 method of \citet{wol67} to the SB2 case along the lines described in \citet{rau00} and \citet{san06a}.}). To finish the analysis of the systems, we estimate, from the spectral classification of each component, the spectroscopic brightness ratio of the system. For that purpose, we compute the mean ratios between the observed equivalent widths (EWs) and the ``canonical'' EWs for stars with same spectral classifications as those of the components of binary systems. These ``canonical'' values are taken from \citet{con71} and \citet{con73}. In addition, we associate these values to EWs measured from synthetic spectra of stars having also similar spectral classifications. The 1-$\sigma$ error-bar given on the brightness ratio corresponds to the dispersions of the ratios measured on each spectral line. This brightness ratio is essential to correct the disentangled spectra in order that they are comparable to those of single stars.

Finally, for all the O-type stars of our sample, we also derive their spectral type. For that, we use the quantitative criteria of \citet{con71}, \citet{con73} and \citet{mat88,mat89}. These criteria rely on the EW ratio of diagnostic lines. These EWs are measured on the highest signal-to-noise spectra of each objects. We adopt, in the present paper, the following usual notations: $\log W' = \log (EW_{4471}) - \log (EW_{4542})$ for the spectral type, $\log W'' = \log (EW_{4089}) - \log (EW_{4143})$ and $\log W''' = \log (EW_{4388}) + \log (EW_{4686})$ for the luminosity class. However, the $\log W''$ criterion is restricted to $\mathrm{O}\,7-\mathrm{O}\,9.7$ stars whilst the $\log W'''$ can only be used for $\mathrm{O}\,8-\mathrm{O}\,9.7$ stars. To be applicable, this last criterion requires that the stars be single or that the brightness ratio between both components be known. The uncertainties on the spectral type determination are generally of about one temperature and luminosity class. To complete this classification, we also use the ``f tag'' notation presented by \citet{wal71} and the subsequent papers.

\newpage
\addtocounter{table}{0}
\longtab{1}{
\begin{longtable}{lcccccc}
\caption{List of stars investigated in this study. $\Delta T$ corresponds to the time elapsed between the first and the last observation of an observing run and is expressed in days, and N represents the number of spectra for each star.}\label{list}\\
\hline\hline
Stars	& Cyg &   Obs. run & Instrument & Wavelength domain & $\Delta T$ & N \\	    
\hline
\endfirsthead
\caption{continued.}\\
\hline\hline
Stars	& Cyg &   Obs. run & Instrument & Wavelength domain & $\Delta T$ & N \\	   
\hline
\endhead
\hline
\endfoot
HD\,193443 & Cyg\,OB1 & Aug. 2004 & OHP - Elodie/1.93m      & $[3850-6850]$\AA & $-$  & 1\\
           &          & Aug. 2005 & OHP - Elodie/1.93m      & $[3850-6850]$\AA & $-$  & 1\\
           &          & Sep. 2008 & OHP - Aur{\'e}lie/1.52m & $[4450-4900]$\AA & 7.06 & 4\\
           &          & Oct. 2008 & OHP - Aur{\'e}lie/1.52m & $[4450-4900]$\AA & 3.17 & 3\\
           &          & Dec. 2009 & OHP - Aur{\'e}lie/1.52m & $[4450-4650]$\AA & 2.98 & 3\\
           &          & Jun. 2010 & OHP - Aur{\'e}lie/1.52m & $[4450-4900]$\AA &19.98 & 13\\
           &          & Aug. 2010 & OHP - Aur{\'e}lie/1.52m & $[4450-4900]$\AA & 5.94 & 7\\
           &          & Dec. 2010 & OHP - Aur{\'e}lie/1.52m & $[4450-4650]$\AA & 4.00 & 5\\
\hline
HD\,193514 & Cyg\,OB1 & Aug. 2004 & OHP - Elodie/1.93m      & $[3850-6850]$\AA & $-$  & 1\\
           &          & Aug. 2005 & OHP - Elodie/1.93m      & $[3850-6850]$\AA & $-$  & 1\\
           &          & Sep. 2008 & OHP - Aur{\'e}lie/1.52m & $[4450-4900]$\AA & 7.17 & 4\\
           &          & Oct. 2008 & OHP - Aur{\'e}lie/1.52m & $[4450-4900]$\AA & 4.07 & 5\\
           &          & Jun. 2009 & SPM - Espresso/2.12m    & $[3780-6950]$\AA & $-$  & 1\\
           &          & Jun. 2010 & OHP - Aur{\'e}lie/1.52m & $[4450-4900]$\AA & 5.83 & 6\\
           &          & Aug. 2010 & OHP - Aur{\'e}lie/1.52m & $[4450-4900]$\AA & 4.93 & 4\\
\hline
HD\,193595 & Cyg\,OB1 & Sep. 2008 & OHP - Aur{\'e}lie/1.52m & $[4450-4900]$\AA & 7.13 & 4\\
           &          & Oct. 2008 & OHP - Aur{\'e}lie/1.52m & $[4450-4900]$\AA & 1.14 & 2\\
           &          & Jun. 2009 & SPM - Espresso/2.12m    & $[3780-6950]$\AA &  $-$ & 1\\
           &          & Aug. 2010 & OHP - Aur{\'e}lie/1.52m & $[4450-4900]$\AA & 4.00 & 2\\
\hline
HD\,193682 & Cyg\,OB1 & Sep. 2008 & OHP - Aur{\'e}lie/1.52m & $[4450-4900]$\AA & 7.02 & 4\\
           &          & Oct. 2008 & OHP - Aur{\'e}lie/1.52m & $[4450-4900]$\AA & 1.06 & 2\\
           &          & Jun. 2009 & SPM - Espresso/2.12m    & $[3780-6950]$\AA &  $-$ & 1\\
           &          & Jun. 2010 & OHP - Aur{\'e}lie/1.52m & $[4450-4900]$\AA &  $-$ & 1\\
           &          & Aug. 2010 & OHP - Aur{\'e}lie/1.52m & $[4450-4900]$\AA & 2.98 & 2\\
\hline
HD\,194094 & Cyg\,OB1 & Jun. 2009 & SPM - Espresso/2.12m    & $[3780-6950]$\AA & 3.93 & 2\\
           &          & Aug. 2010 & OHP - Aur{\'e}lie/1.52m & $[4450-4900]$\AA & 4.94 & 2\\
           &          & Sep. 2011 & SPM - Espresso/2.12m    & $[3780-6950]$\AA &  $-$ & 1\\
\hline
HD\,194280 & Cyg\,OB1 & Oct. 2004 & OHP - Aur{\'e}lie/1.52m & $[4450-4900]$\AA & 4.05 & 3\\
           &          & Jun. 2005 & OHP - Aur{\'e}lie/1.52m & $[4450-4900]$\AA & 8.03 & 7\\
           &          & Sep. 2006 & OHP - Aur{\'e}lie/1.52m & $[4450-4900]$\AA & 4.07 & 8\\
           &          & Oct. 2006 & OHP - Aur{\'e}lie/1.52m & $[4450-4900]$\AA & $-$  & 1\\
           &          & Nov. 2007 & OHP - Aur{\'e}lie/1.52m & $[4450-4900]$\AA & 26.95& 23\\
\hline
HD\,228841 & Cyg\,OB1 & Sep. 2008 & OHP - Aur{\'e}lie/1.52m & $[4450-4900]$\AA & 6.08 & 3\\
           &          & Oct. 2008 & OHP - Aur{\'e}lie/1.52m & $[4450-4900]$\AA & $-$  & 1\\
           &          & Jun. 2009 & SPM - Espresso/2.12m    & $[3780-6950]$\AA & $-$  & 1\\
           &          & Jun. 2010 & OHP - Aur{\'e}lie/1.52m & $[4450-4900]$\AA &  $-$ & 1\\
           &          & Aug. 2010 & OHP - Aur{\'e}lie/1.52m & $[4450-4900]$\AA & 3.18 & 2\\
\hline
HD\,228989 & Cyg\,OB1 & Aug. 2009 & SPM - Espresso/2.12m    & $[3780-6950]$\AA & 0.80 & 2\\
           &          & Jul. 2010 & SPM - Espresso/2.12m    & $[3780-6950]$\AA & $-$  & 1\\
           &          & Aug. 2010 & OHP - Aur{\'e}lie/1.52m & $[4450-4900]$\AA & 5.13 & 5\\
           &          & Jun. 2011 & SPM - Espresso/2.12m    & $[3780-6950]$\AA & 3.94 & 5\\
           &          & Sep. 2011 & SPM - Espresso/2.12m    & $[3780-6950]$\AA & 1.98 & 5\\
\hline
HD\,229234 & Cyg\,OB1 & Sep. 2008 & OHP - Aur{\'e}lie/1.52m & $[4450-4900]$\AA & 1.06 & 2\\
           &          & Oct. 2008 & OHP - Aur{\'e}lie/1.52m & $[4450-4900]$\AA & 1.82 & 2\\
           &          & Jun. 2009 & SPM - Espresso/2.12m    & $[3780-6950]$\AA & $-$  & 1\\
           &          & Aug. 2009 & SPM - Espresso/2.12m    & $[3780-6950]$\AA & 0.96 & 2\\
           &          & Jun. 2010 & OHP - Aur{\'e}lie/1.52m & $[4450-4900]$\AA & 6.06 & 4\\
           &          & Jul. 2010 & SPM - Espresso/2.12m    & $[3780-6950]$\AA & 1.06 & 2\\
           &          & Aug. 2010 & OHP - Aur{\'e}lie/1.52m & $[4450-4900]$\AA & 5.86 & 6\\
\hline
HD\,190864 & Cyg\,OB3 & Aug. 2001 & OHP - Elodie/1.93m      & $[3850-6850]$\AA & $-$  & 1\\
           &          & Sep. 2008 & OHP - Aur{\'e}lie/1.52m & $[4450-4900]$\AA & 7.05 & 5\\
           &          & Oct. 2008 & OHP - Aur{\'e}lie/1.52m & $[4450-4900]$\AA & 3.25 & 3\\
           &          & Jun. 2009 & SPM - Espresso/2.12m    & $[3780-6950]$\AA & 5.91 & 2\\
           &          & Jun. 2010 & OHP - Aur{\'e}lie/1.52m & $[4450-4900]$\AA & 1.00 & 2\\
           &          & Aug. 2010 & OHP - Aur{\'e}lie/1.52m & $[4450-4900]$\AA & 4.00 & 4\\
\hline
HD\,227018 & Cyg\,OB3 & Sep. 2008 & OHP - Aur{\'e}lie/1.52m & $[4450-4900]$\AA & 7.05 & 4\\
           &          & Oct. 2008 & OHP - Aur{\'e}lie/1.52m & $[4450-4900]$\AA & 4.02 & 4\\
           &          & Aug. 2010 & OHP - Aur{\'e}lie/1.52m & $[4450-4900]$\AA & 2.02 & 2\\
\hline
HD\,227245 & Cyg\,OB3 & Aug. 2009 & SPM - Espresso/2.12m    & $[3780-6950]$\AA & 0.86 & 2\\
           &          & Jun. 2011 & SPM - Espresso/2.12m    & $[3780-6950]$\AA & 3.81 & 3\\
\hline
HD\,227757 & Cyg\,OB3 & Jun. 2009 & SPM - Espresso/2.12m    & $[3780-6950]$\AA & 4.98 & 2\\
           &          & Jul. 2010 & SPM - Espresso/2.12m    & $[3780-6950]$\AA & $-$  & 1\\
           &          & Jun. 2011 & SPM - Espresso/2.12m    & $[3780-6950]$\AA & 3.94 & 3\\
\hline
HD\,191423 & Cyg\,OB8 & Aug. 2004 & OHP - Elodie/1.93m      & $[3850-6850]$\AA & $-$  & 1\\
           &          & Sep. 2008 & OHP - Aur{\'e}lie/1.52m & $[4450-4900]$\AA & 7.09 & 4\\
           &          & Oct. 2008 & OHP - Aur{\'e}lie/1.52m & $[4450-4900]$\AA & $-$  & 1\\
           &          & Jun. 2009 & SPM - Espresso/2.12m    & $[3780-6950]$\AA & $-$  & 1\\
           &          & Aug. 2010 & OHP - Aur{\'e}lie/1.52m & $[4450-4900]$\AA & 3.17 & 2\\
\hline
HD\,191978 & Cyg\,OB8 & Aug. 2001 & OHP - Elodie/1.93m      & $[3850-6850]$\AA & 1.05 & 2\\
           &          & Sep. 2008 & OHP - Aur{\'e}lie/1.52m & $[4450-4900]$\AA & 5.94 & 3\\
           &          & Oct. 2008 & OHP - Aur{\'e}lie/1.52m & $[4450-4900]$\AA & $-$  & 1\\
           &          & Jun. 2009 & SPM - Espresso/2.12m    & $[3780-6950]$\AA & 8.92 & 2\\
           &          & Aug. 2010 & OHP - Aur{\'e}lie/1.52m & $[4450-4900]$\AA & 4.11 & 3\\
\hline 
HD\,193117 & Cyg\,OB8 & Sep. 2008 & OHP - Aur{\'e}lie/1.52m & $[4450-4900]$\AA & 6.82 & 3\\
           &          & Oct. 2008 & OHP - Aur{\'e}lie/1.52m & $[4450-4900]$\AA & $-$  & 1\\
           &          & Jun. 2009 & SPM - Espresso/2.12m    & $[3780-6950]$\AA & 5.05 & 2\\
           &          & Aug. 2010 & OHP - Aur{\'e}lie/1.52m & $[4450-4900]$\AA & 3.20 & 2\\
\hline
HD\,194334 & Cyg\,OB9 & Sep. 2008 & OHP - Aur{\'e}lie/1.52m & $[4450-4900]$\AA & 6.75 & 2\\
           &          & Oct. 2008 & OHP - Aur{\'e}lie/1.52m & $[4450-4900]$\AA & $-$  & 1\\
           &          & Jun. 2009 & SPM - Espresso/2.12m    & $[3780-6950]$\AA & $-$  & 1\\
           &          & Aug. 2009 & SPM - Espresso/2.12m    & $[3780-6950]$\AA & 0.87 & 2\\
           &          & Jun. 2010 & OHP - Aur{\'e}lie/1.52m & $[4450-4900]$\AA & 1.07 & 2\\
           &          & Jul. 2010 & SPM - Espresso/2.12m    & $[3780-6950]$\AA & 1.03 & 2\\
           &          & Aug. 2010 & OHP - Aur{\'e}lie/1.52m & $[4450-4900]$\AA & 3.87 & 1\\
\hline
HD\,194649 & Cyg\,OB9 & Jun. 2009 & SPM - Espresso/2.12m    & $[3780-6950]$\AA & $-$  & 1\\
           &          & Aug. 2009 & SPM - Espresso/2.12m    & $[3780-6950]$\AA & $-$  & 1\\
           &          & Jun. 2010 & OHP - Aur{\'e}lie/1.52m & $[4450-4900]$\AA & 5.95 & 3\\
           &          & Aug. 2010 & OHP - Aur{\'e}lie/1.52m & $[4450-4900]$\AA & 2.13 & 2\\
           &          & Jun. 2011 & SPM - Espresso/2.12m    & $[3780-6950]$\AA & 3.93 & 5\\
           &          & Sep. 2011 & SPM - Espresso/2.12m    & $[3780-6950]$\AA & 3.98 & 5\\
\hline
HD\,195213 & Cyg\,OB9 & Sep. 2008 & OHP - Aur{\'e}lie/1.52m & $[4450-4900]$\AA & 6.96 & 3\\
           &          & Oct. 2008 & OHP - Aur{\'e}lie/1.52m & $[4450-4900]$\AA & $-$  & 1\\
           &          & Aug. 2009 & SPM - Espresso/2.12m    & $[3780-6950]$\AA & 0.96 & 2\\
           &          & Jun. 2010 & OHP - Aur{\'e}lie/1.52m & $[4450-4900]$\AA & $-$  & 1\\
           &          & Jul. 2010 & SPM - Espresso/2.12m    & $[3780-6950]$\AA & $-$  & 1\\
           &          & Aug. 2010 & OHP - Aur{\'e}lie/1.52m & $[4450-4900]$\AA & 4.91 & 2\\
\hline
\end{longtable}
}

\newpage
\begin{sidewaystable*}[htbp]
\begin{center}
\caption{Journal of the spectroscopic observations of the O-type stars. The first column gives the HJD at mid-exposure. The following columns provide, for each spectral line, the measured RVs (in \kms). The full table is available at the CDS.}\label{rv}
\begin{tabular}{lcccccccccc}
\hline\hline
\multicolumn{11}{c}{HD\,190864}\\
\hline
HJD  & \ion{He}{i}~4471 &  \ion{He}{ii}~4542  &  \ion{N}{iii}~4634 & \ion{N}{iii}~4641 &  \ion{He}{ii}~4686  & \ion{He}{i}~4713 & \ion{He}{i}~5876 & \ion{Na}{i}~5889 & \ion{Na}{i}~5895  &  Disent\\
$-$ 2\,450\,000 &  [\kms]  &  [\kms] & [\kms]  &  [\kms]  & [\kms] &  [\kms]  &  [\kms]  &  [\kms] &   [\kms] &  [\kms]\\
\hline
2134.4733  &  $-$13.0   &    $-$5.2  &    $-$19.3  &   $-$26.3   &     $-$4.9  &   $-$18.6  & $-$13.9 & $-$19.1 &  $-$18.3&$-$\\
4711.3132  &  $-$16.4   &    $-$7.7  &    $-$23.4  &   $-$29.5   &     11.6  &   $-$20.9  &  $-$  &  $-$  &  $-$	&$-$\\
4712.3000  &  $-$15.5   &    $-$4.3  &    $-$20.5  &   $-$24.8   &     13.4  &   $-$11.9  &  $-$  &  $-$  &  $-$	&$-$\\
4712.3111  &  $-$13.7   &    $-$7.2  &    $-$30.3  &   $-$24.2   &     18.0  &   $-$15.7  &  $-$  &  $-$  &  $-$	&$-$\\
4717.2954  &  $-$16.0   &    $-$6.9  &    $-$18.9  &   $-$28.4   &     17.1  &   $-$16.6   & $-$  &  $-$  &  $-$	&$-$\\
4718.3637  &  $-$15.9   &    $-$3.9  &    $-$19.7  &   $-$35.9   &     13.4  &   $-$14.2  &  $-$  &  $-$  &  $-$	&$-$\\
4740.2629  &  $-$15.8    &   $-$8.5  &    $-$20.5  &   $-$30.9   &     10.9  &   $-$18.3  &  $-$  &  $-$  &  $-$	&$-$\\
4741.3195  &  $-$11.0    &   $-$5.1  &    $-$18.1 &    $-$28.3   &     11.7  &    $-$9.6  &  $-$  &  $-$  &  $-$	&$-$\\
4743.5116  &  $-$12.2    &   $-$6.5  &    $-$15.7 &    $-$28.2   &     13.0  &   $-$13.2  &  $-$  &  $-$  &  $-$	&$-$\\
4982.8779  &  $-$20.7    &   $-$9.2  &    $-$28.3 &    $-$62.0   &     $-$3.5  &   $-$25.4  &   $-$13.3  &   $-$20.1  &   $-$18.6	&$-$\\
4988.7847  &  $-$16.1    &   $-$0.8  &    $-$10.6 &    $-$32.4   &    $-$27.7  &    $-$   &   $-$19.7  &   $-$18.4  &   $-$18.5	&$-$\\
5355.4081  &  $-$17.0    &   $-$4.0  &    $-$16.2 &    $-$25.1   &      1.0  &   $-$16.1  &  $-$  &  $-$  &  $-$	&$-$\\
5356.4053  &  $-$14.3    &   $-$7.2  &    $-$17.3 &    $-$34.0   &     $-$1.2  &   $-$14.1  &  $-$  &  $-$  &  $-$	&$-$\\
5415.3437  &  $-$14.0    &   $-$5.2	 &  $-$21.6   &  $-$29.5     &    4.5    &   $-$14.8  &  $-$  &  $-$  &  $-$	&$-$\\
5416.3784  &  $-$18.8    &  $-$10.9  &    $-$25.0 &    $-$33.6   &     $-$7.0  &   $-$20.2  &  $-$  &  $-$  &  $-$	&$-$\\
5417.3323  &  $-$18.5    &   $-$8.2  &    $-$24.3 &    $-$34.4   &      7.6  &   $-$16.9  &  $-$  &  $-$  &  $-$	&$-$\\
5419.3479  &  $-$13.8    &   $-$7.1  &    $-$20.4 &    $-$30.7   &      0.4  &   $-$16.9  &  $-$  &  $-$  &  $-$	&$-$\\
\hline
\multicolumn{11}{c}{HD\,227018}\\
\hline
4711.3323  &  12.7   &   23.3   & $-$  & $-$      & 24.4  &   15.0  & $-$ &  $-$  & $-$ &$-$ \\
4712.3346  &  20.2   &   23.8	& $-$  & $-$	& 27.6  &   21.1  & $-$ &  $-$  & $-$ &$-$ \\
4717.3188  &  16.0   &   20.9	& $-$  & $-$	& 25.3  &   15.0  & $-$ &  $-$  & $-$ &$-$ \\
4718.3862  &  16.4   &   22.5	& $-$  & $-$	& 23.6  &   13.6  & $-$ &  $-$  & $-$ &$-$ \\
4740.2871  &  12.1   &   15.3	& $-$  & $-$	& 20.6  &   13.4  & $-$ &  $-$  & $-$ &$-$ \\
4741.3477  &  12.0   &   15.8	& $-$  & $-$	& 16.8  &  -----  & $-$ &  $-$  & $-$ &$-$ \\
4742.2887  &  11.9   &   19.5	& $-$  & $-$	& 22.7  &   16.4  & $-$ &  $-$  & $-$ &$-$ \\
4744.3032  &  15.6   &   21.6	& $-$  & $-$	& 22.8  &   13.0  & $-$ &  $-$  & $-$ &$-$ \\
5417.3604  &  18.2   &   26.2	& $-$  & $-$	& 27.1	&   17.0  & $-$ &  $-$  & $-$ &$-$ \\
5419.3755  &  20.8   &   27.7	& $-$  & $-$	& 26.7	&   24.5  & $-$ &  $-$  & $-$ &$-$ \\
\hline
\end{tabular}
\end{center}
\end{sidewaystable*}

\section{O-type stars in the Cyg\,OB1 association}
\label{sect:ob1}
\subsection{Gravitationally bound systems}
\label{subsec:bound1}
\subsubsection{HD\,193443}
\begin{figure}[htbp]
\begin{center}
\includegraphics[width=7cm,bb=32 164 579 704,clip]{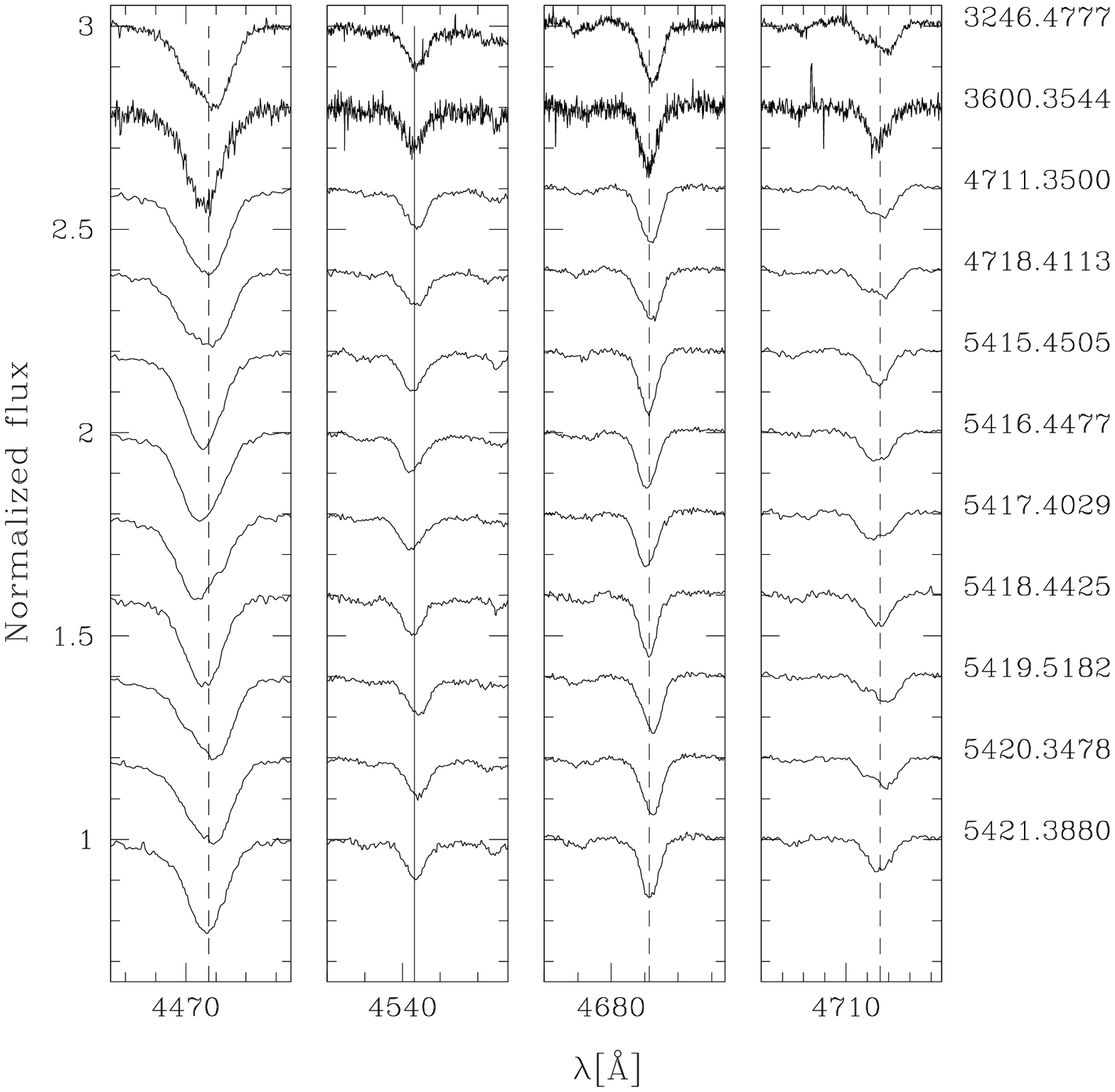}
\includegraphics[width=7cm,bb=32 164 555 676,clip]{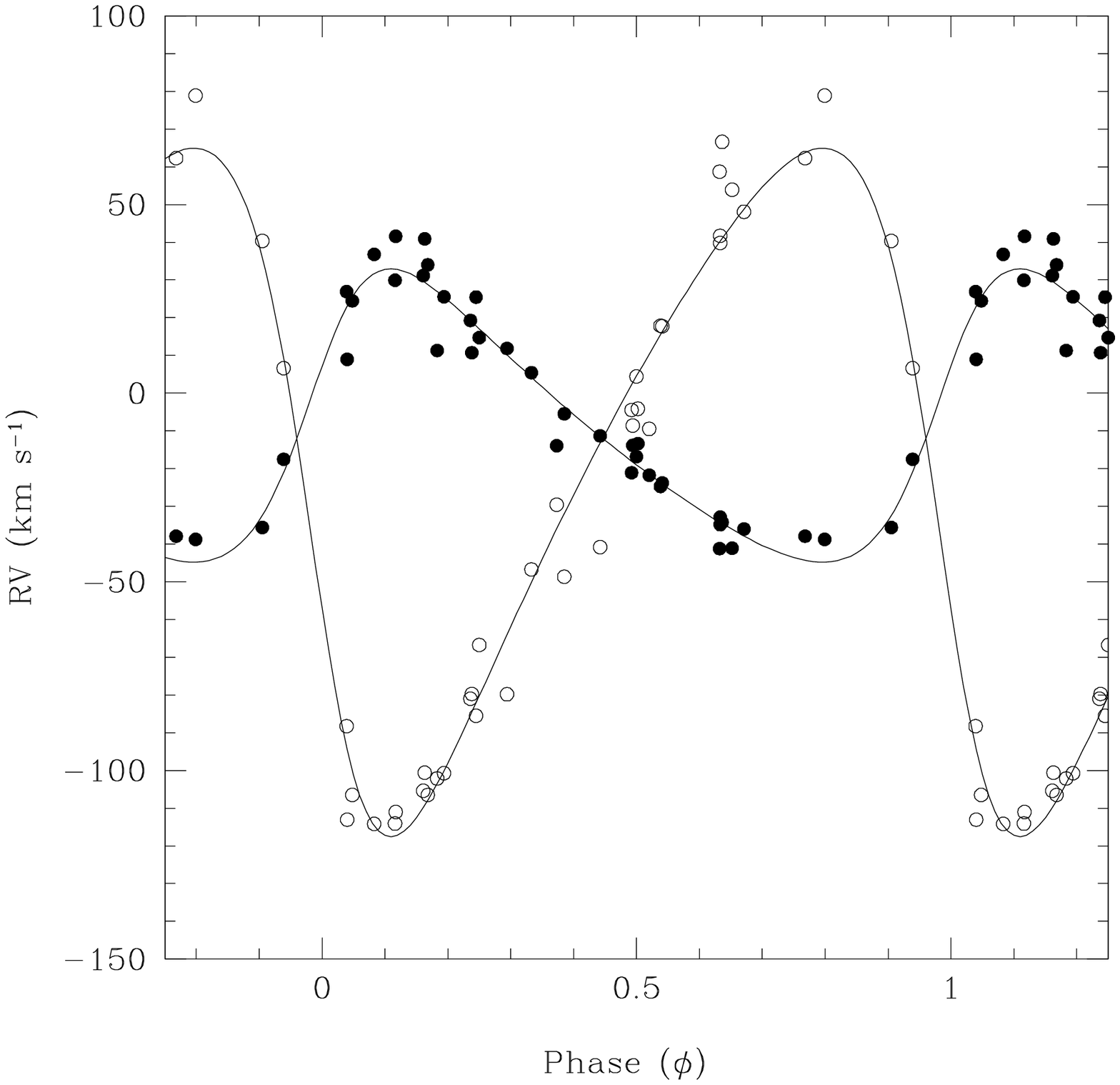}
\caption{{\it Top:} \ion{He}{i}~4471, \ion{He}{ii}~4542, \ion{He}{ii}~4686, \ion{He}{i}~4713 line profiles of HD\,193443 at different epochs. {\it Bottom:} RV curves of HD\,193443 computed for a period of 7.467~days. Filled circles represent the primary RVs whilst the open circles correspond to the secondary. }\label{sol_hd193443}
\end{center}
\end{figure}

Reported as binary by \citet{mul54}, this star has never been the target of an intense spectroscopic monitoring. Therefore, no orbital solution is known for the system nor are there spectral classifications for the components. We acquired thirty-seven spectra of HD\,193443 from Aug 2004 ($\mathrm{HJD} = 3246.4777$) to Dec 2010 ($\mathrm{HJD} = 5543.2387$). These data reveal a periodic variation of the line profiles, clearly visible in the \ion{He}{i}~4471 and \ion{He}{i}~4713 lines. Even though the signature of the secondary is detected, the small RV separation between both components prevents us from completely resolving the individual line profiles throughout the orbit. The secondary's lines are thus never totally separated from those of the primary even for the helium lines or the metallic lines as it is shown in the upper panel of Fig.~\ref{sol_hd193443}.

The analysis of the SB2 system is performed as explained in Section~\ref{sect:obs}. However, the Gaussian fit is made on the bottom of the absorption lines of the primary component. At the maximum of the RV separation, we also fit a second profile to determine the RVs of the secondary. The HMM technique measured on the RVs refined by the disentangling programme yields a period of $7.467 \pm 0.003$ days. From this period, we compute the orbital parameters of HD\,193443 listed in Table~\ref{sol_orb} and the best fit of the RV curves, shown in Fig.~\ref{sol_hd193443}, is obtained for a non-zero eccentricity.

\begin{figure}[htbp]
\begin{center}
\includegraphics[width=8.5cm,bb=31 422 569 695,clip]{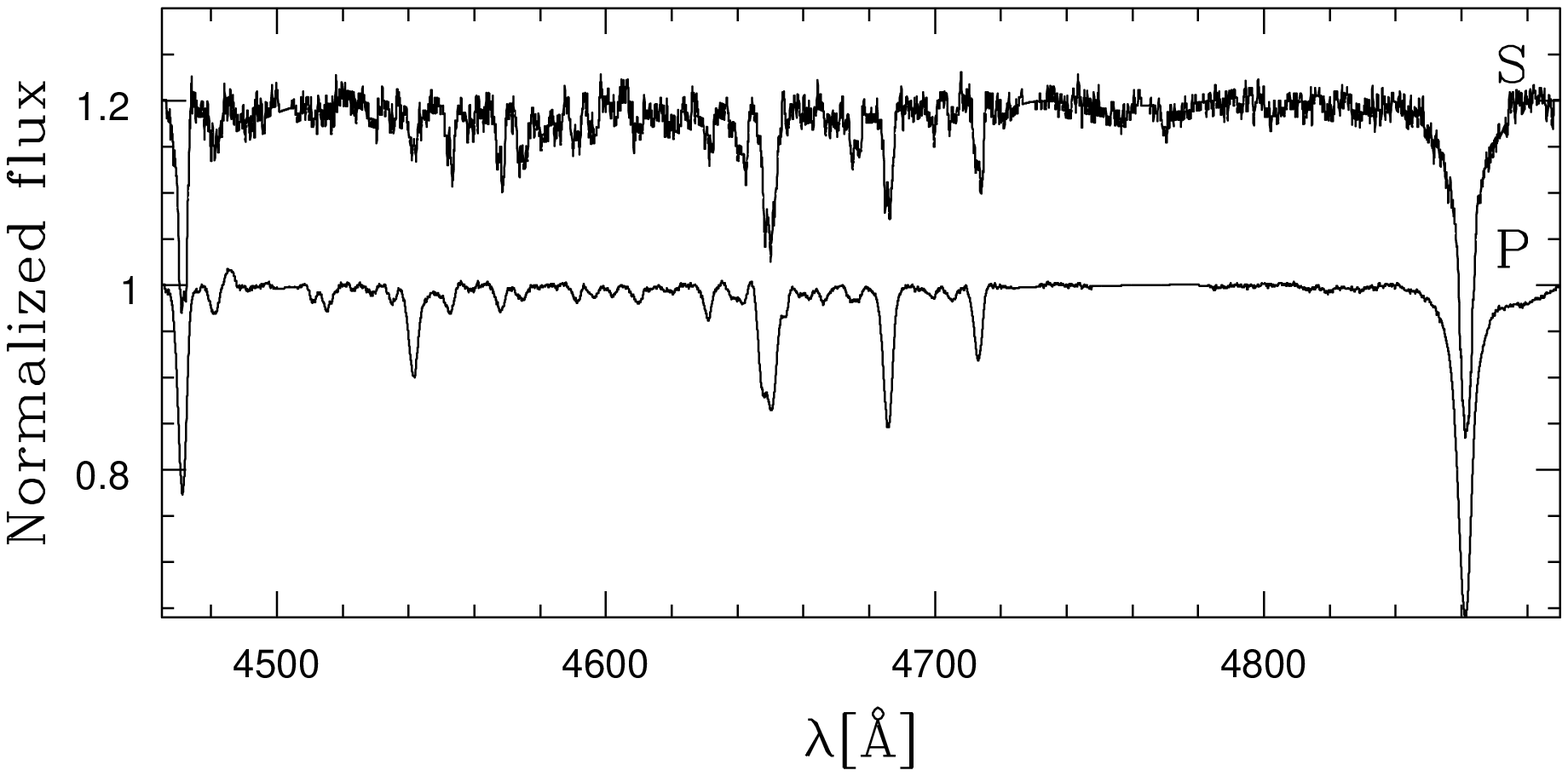}
\caption{Disentangled spectra of the two components of HD\,193443. The spectra are normalized accounting for the optical brightness ratio of 3.9. The secondary spectrum is vertically shifted for clarity.}\label{spectra_hd193443}
\end{center}
\end{figure}

The study of the disentangled spectra of HD\,193443 shows that the \ion{He}{ii}~4542 line is present but is weak for both components relative to the \ion{He}{i}~4471 line. This  indicates two late O-type stars. We compute, from the observed spectra, $\log W' = 0.321$ and $\log W'=0.483$ for the primary and the secondary, respectively, thereby corresponding to respective subtypes of O\,9 and O\,9.5. However, our dataset does not allow us to constrain the luminosity classes of these stars. Indeed, the majority of our data does not cover the \ion{Si}{iv}~4089 and \ion{He}{i}~4143 lines. Moreover, for those that do, the RV separation is insufficient to resolve with accuracy the signature of the secondary component and thus to correctly fit it with Gaussian profiles. To determine the brightness ratio, we measure the EWs of the \ion{He}{i}~4471, 4713 and \ion{He}{ii}~4542 lines of the disentangled and observed spectra at the maximum of separation and we compare these values with those of \citet{con71} for stars of the same spectral types as both components. Moreover, we also measure the EWs of the above-quoted lines on synthetic spectra with the same spectral classifications as explained in Section~\ref{sect:obs}. Even though the luminosity class is unknown for both components of HD\,193443, the lines that we selected for this computation are only slightly affected by the luminosity class of the stars. Therefore, we decide to take into account the ``canonical'' EWs of all the stars with only similar subtypes. The ratios of these values provide a brightness ratio of about $3.9 \pm 0.4$ between the primary and the secondary components. The disentangled spectra, corrected from the brightness ratio, are shown in Fig.~\ref{spectra_hd193443}.

The properties of both components thus show two stars with relatively close spectral types but with large mass and brightness ratios. This indicates that the primary is rather evolved relative to the secondary. First, we assume that HD\,193443 is either an O\,9III$+$O\,9.5V or an O\,9I$+$O\,9.5III system. In the former case, the comparison between the minimum masses and the values quoted in \citet[][$22$~\msun\ and $16$~\msun\ for the primary and the secondary, respectively]{mar05} suggests an inclination between $17^{\circ}$ and $21^{\circ}$. In the latter case, we obtain, from masses estimated to $30$~\msun\ for the primary and $21$~\msun\ for the secondary, an inclination between $15^{\circ}$ and $19^{\circ}$. Therefore, we consider that the inclination of the system is very low and probably ranges between $15^{\circ}$ and $22^{\circ}$. From the formula of \citet{egg83}, we compute the projected Roche lobe radii ($RRL \sin i$) of 8.3~\rsun\ and $5.6$~\rsun\ for the primary and the secondary components. These values indicate Roche lobe radii between $22$~\rsun\ and $32$~\rsun\ for the primary and between $15$~\rsun\ and $22$~\rsun\ for the secondary. In both cases, the secondary (which would have a radius of $7$~\rsun\ or $13$~\rsun\ depending on the case) and the primary (with a radius between 13~\rsun\ and 22~\rsun) would not fill their Roche lobe. Secondly, if the stellar classifications were O\,9I for the primary and O\,9.5V for the secondary, the values of the inclination would be estimated between $15^{\circ}$ and $21^{\circ}$, thereby implying Roche lobe radii between $23$~\rsun\ and $32$~\rsun\ for the primary and $16$~\rsun\ and $22$~\rsun\ for the secondary. In this case, the primary could fill its Roche lobe. At this stage and for the other SB2 systems, it is important to note that the analysis of the orbital parameters and the estimation of the inclination of the SB2 systems are only made to give a first idea of the geometry of the systems. However, we must be careful on the information provided here. Indeed, the masses given in the tables of \citet{mar05} are standard values that also depend on other parameters such as e.g., \logg, \teff\ and/or $\log(L/L_{\odot})$. The values quoted in these tables can thus be different from the real parameters of both components. Therefore, it is common that the mass ratios computed from these standard values do not correspond to the real mass ratios determined from the orbital solutions.

\subsubsection{HD\,228989}

Located in the Berkeley\,86 young open cluster \citep{for81}, this star was quoted as an O\,9.5$+$O\,9.5 binary system by \citet{mas95}. However, no orbital solution of this object was yet determined.

\begin{figure}[htbp]
\begin{center}
\includegraphics[width=7cm,bb=32 164 579 704,clip]{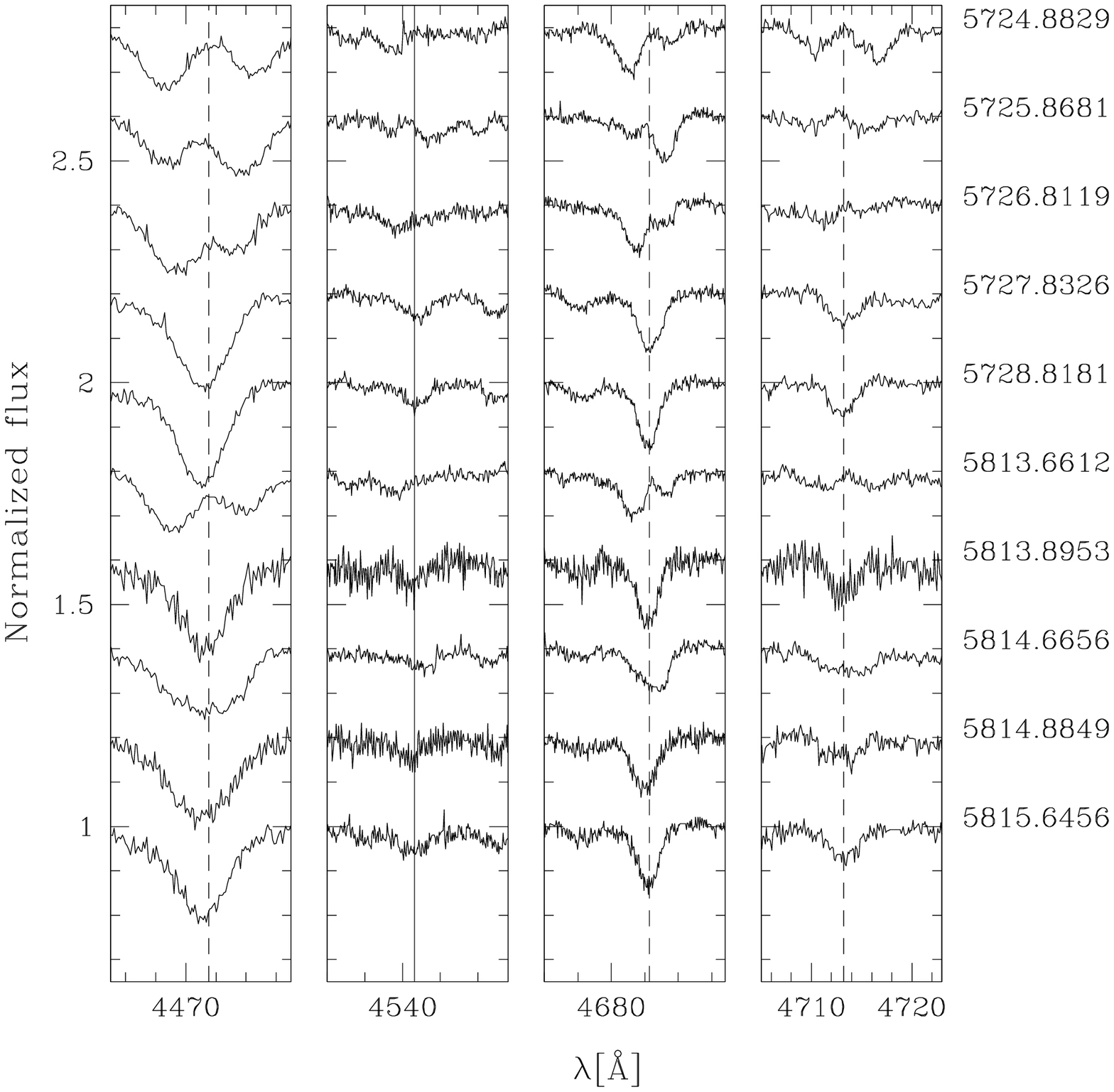}
\includegraphics[width=7cm,bb=32 164 555 676,clip]{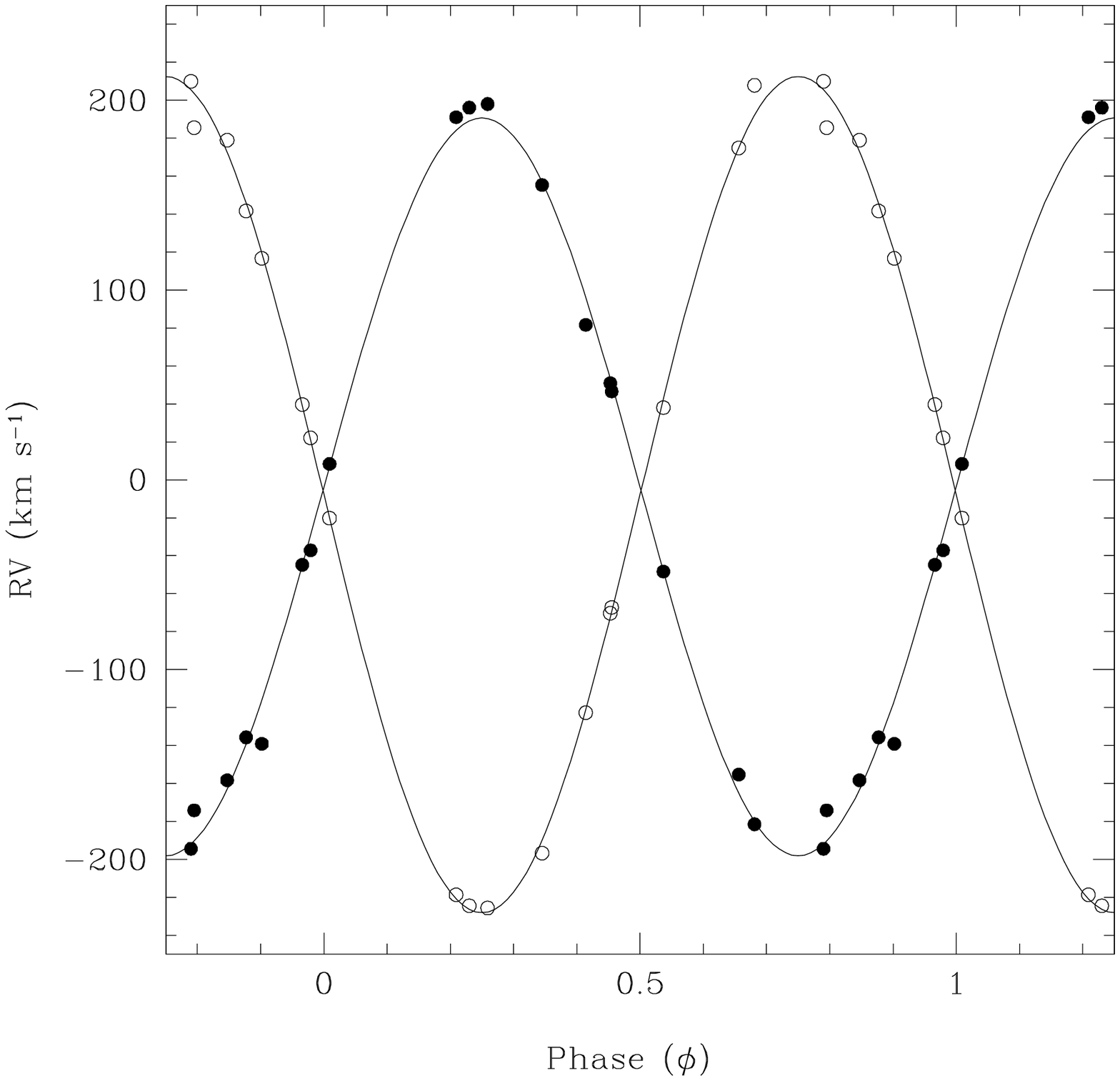}
\caption{{\it Top:} \ion{He}{i}~4471, \ion{He}{ii}~4542, \ion{He}{ii}~4686, \ion{He}{i}~4713 line profiles at different epochs. {\it Bottom:} RV curves of HD\,228989 computed from $P_{\mathrm{orb}} = 1.77352$~days. Filled circles represent the primary whilst the open circles correspond to the secondary.}\label{sol_hd228989}
\end{center}
\end{figure}

\begin{figure}[htbp]
\begin{center}
\includegraphics[width=8.5cm,bb=20 420 579 695,clip]{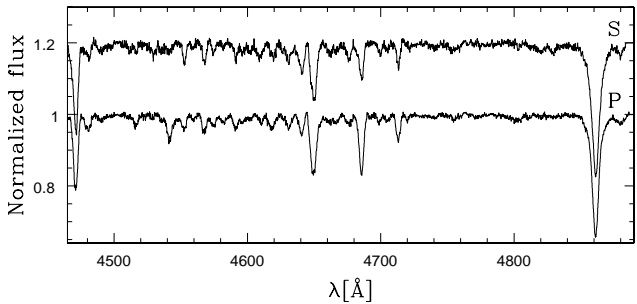}
\caption{Disentangled spectra of the two components of HD\,228989. The spectra are normalized accounting for a brightness ratio of 1.2. The secondary spectrum is vertically shifted for clarity.}\label{spectra_hd228989}
\end{center}
\end{figure}

We collected eighteen observations of this object between Aug 2009 ($\mathrm{HJD}=5049.9519$) and Sep 2011 ($\mathrm{HJD}=5815.6460$). These data clearly reveal, as it is shown in the upper panel of Fig.~\ref{sol_hd228989}, a binary system composed of two stars with rather similar spectral classifications. Our analysis of the refined RVs gives an orbital period of 1.77352~days. We compute from this value the orbital parameters reported in Table~\ref{sol_orb}. We have computed two orbital solutions: an eccentric and a circular. The rms being smaller for the circular orbit, we decide to fix the eccentricity to zero. The RV curves computed with LOSP are shown in the lower panel of Fig.~\ref{sol_hd228989}. As suggested from the disentangled spectra and the mass ratio, the two components seem to be quite close to each other in terms of the physical properties. 
From the raw (i.e., not corrected by the brightness ratio) disentangled spectra and the observed spectra, we determine spectral types of O\,8.5 ($\log W' =0.238$) and O\,9.7 ($\log W' = 0.791$) for the primary and the secondary components, respectively. In addition, we obtain, on the most deblended spectra, for the primary $\log W''=0.027$ and for the secondary $\log W''=0.104$, suggesting a main-sequence luminosity class for both components. Knowing the spectral classifications of these two stars, we estimate a brightness ratio of about $1.2\pm0.1$, the primary being slightly brighter than its companion. This brightness ratio coupled with the luminosities listed in tables of \citet{mar05} seem to agree with the luminosity classes of both components. Finally, we display in Fig.~\ref{spectra_hd228989} the disentangled spectra corrected for the relative brightness.

The comparison between the minimum masses and the masses published by \citet[][$19$~\msun\ and $<17$~\msun\ for the primary and the secondary, respectively]{mar05} suggests that the system would have an inclination between $42^{\circ}$ and $50^{\circ}$. We stress however that the O\,9.7 subtype is not included in the table of \citet{mar05}. Therefore, we consider the values of the O\,9.5 subtype as an upper limit. On the basis of Eggleton's formula (1983), we estimate $RRL \sin i$ of about $5.7$~\rsun\ and $5.3$~\rsun\ for the primary and the secondary components. From the inclination range estimated from the masses, we compute a Roche lobe radius between $7.4$~\rsun\ and $8.4$~\rsun\ for the primary and between $7.0$~\rsun\ and $8.0$~\rsun\ for the secondary. Such values seem to indicate that both components are very close to fill their Roche lobe (standard radii are of 7.9~\rsun\ and $< 7.2$~\rsun\ for the primary and the secondary, respectively). 

\subsubsection{HD\,229234}

Quoted as belonging to NGC\,6913 \citep{hum78}, HD\,229234 was classified as an O\,9If star by \citet{mas95} whilst \citet{neg04} reported an O\,9II classification and \citet{mor55} reported this star as O\,9.5III. Previous investigations of \citet{liu89} and \citet{boe04} proposed this star to be a binary system with an orbital period of 3.5105 days. In addition, \citet{boe04} published the first SB1 orbital solution for this star. This SB1 binary status was moreover confirmed by a recent analysis of \citet{malchenko2009}. 

\begin{figure}[htbp]
\begin{center}
\includegraphics[width=8.5cm,bb=20 180 575 701,clip]{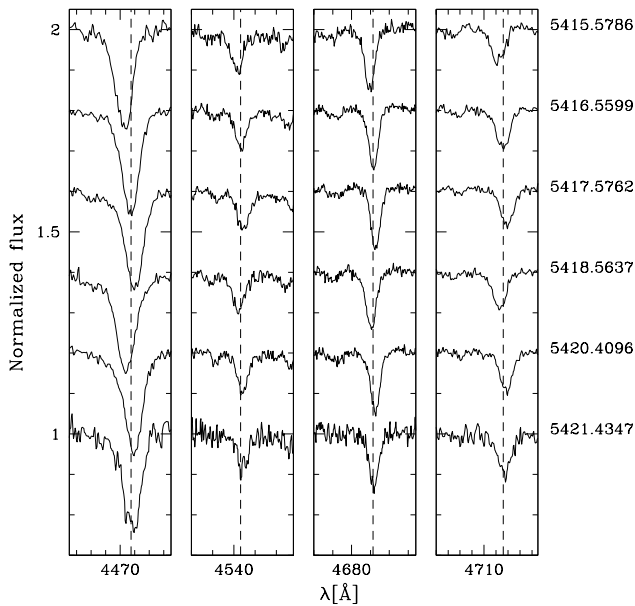}
\caption{Short-term variations of the \ion{He}{i}~4471, \ion{He}{ii}~4542, \ion{He}{ii}~4686 and \ion{He}{i}~4713 line profiles of HD\,229234. The rest wavelengths are given by the dashed lines.}\label{line_hd229234}
\end{center}
\end{figure}

\begin{figure}[htbp]
\begin{center}
\includegraphics[width=7cm,bb=30 173 550 675,clip]{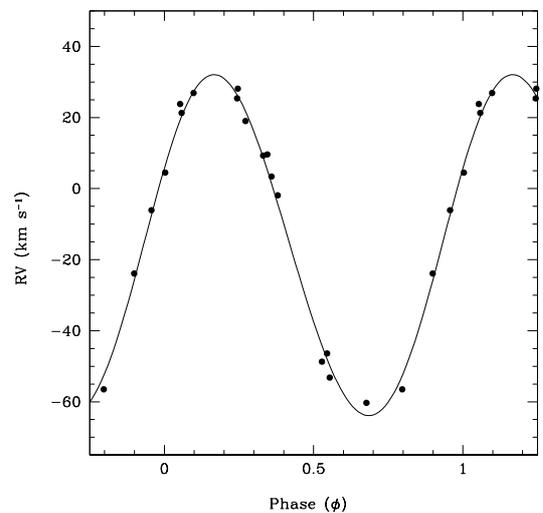}
\caption{RV curve of HD\,229234 computed from P$_{\mathrm{orb}} = 3.510595$~days.}\label{sol_orb_hd229234}
\end{center}
\end{figure}

We obtained nineteen spectra of HD\,229234 between Sep 2008 ($\mathrm{HJD}=4717.3892$) and Aug 2010 ($\mathrm{HJD} = 5421.4347$). This object obviously displays significant variations in its RVs on a short timescale. Fig.~\ref{line_hd229234} shows the Doppler shifts of several spectral lines for six consecutive nights. The periodic motion is clearly visible for all the spectral lines. The mean radial velocities ($\overline{\mathrm{RVs}}$)~$\pm$~the standard deviation computed from the RVs are $-8.7\pm 33.8$, $4.9\pm34.4$, $1.7\pm36.5$ and $-6.5\pm34.0$~\kms\ for the \ion{He}{i}~4471, \ion{He}{ii}~4542, \ion{He}{ii}~4686 and \ion{He}{i}~4713 lines, respectively. No SB2 signature was found in the optical data. We determine on the basis of the HMM technique an orbital period of $3.510595 \pm 0.000245$~days, agreeing with the previous estimates of \citet{liu89} and \citet{boe04}. The parameters of the SB1 orbital solution are listed in Table~\ref{sol_orb} and the RV curve is displayed in Fig.~\ref{sol_orb_hd229234}. These parameters are globally close to those obtained by \citet{boe04}. With the absence of uncertainties in the orbital solution given by these authors, the comparison remains however difficult. The main difference is regardless observed in $T_0$ value because of a different definition of the phase $\phi = 0.0$.

We also derive, for the main component of this SB1 system, an O\,9III classification from the different EW ratios $\log W'= 0.372$, $\log W''= 0.280$ and $\log W'''= 5.325$. This spectral type appears to be in agreement with the different classifications found in the literature. Unlike \citet{mas95}, our classification is not only based on a visual comparison with spectra of the atlas of \citet{Walborn1990}, but also relies on quantitative values.

From the mass of the primary \citep[$22.0 \pm 3.0$~\msun][]{mar05} and the $f_{\mathrm{mass}}$ value given in Table~\ref{sol_orb}, we determine, as minimum mass for the secondary component of the system, a value of about $3.0 \pm 0.3$ \msun.

\subsection{Presumably single stars}
\label{subsec:single1}

\subsubsection{HD\,193514}
This star was classified as O\,7Ib(f) by \citet{rep04}. These authors also reported on a weak emission in the \ion{He}{ii}~4686 line profile. Changes from day to day were already observed by \citet{und95} in the H$\alpha$ profile. However, none of these authors did attribute these variations to a binary nature. 

\begin{figure}[htbp]
\begin{center}
\includegraphics[width=8.5cm,bb=18 160 584 701,clip]{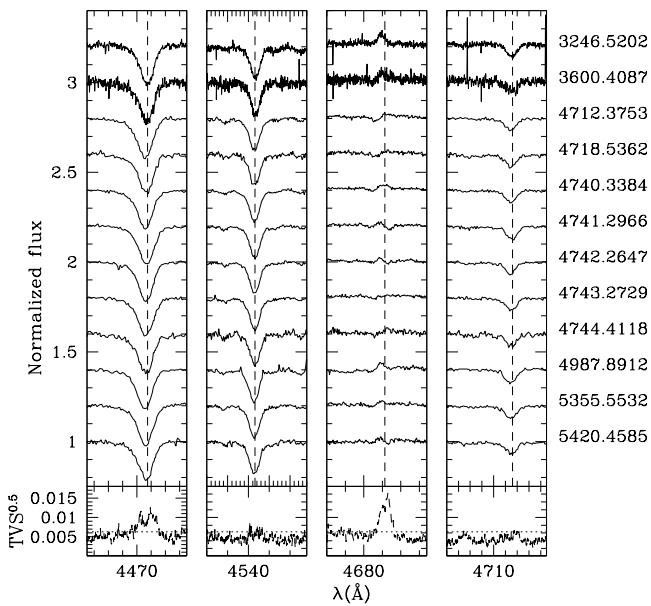}
\caption{The \ion{He}{i}~4471, \ion{He}{ii}~4542, \ion{He}{ii}~4686 and \ion{He}{i}~4713 line profiles of HD\,193514 according to HJD. The rest wavelengths are given by the dashed lines. At the bottom, TVS of each spectral line computed from the Aur{\'e}lie data. The dotted line illustrates the 1\% significance level for the variability evaluated following the approach of \citet{ful96}.}\label{prof_hd193514}
\end{center}
\end{figure}

The twenty-two spectra of HD\,193514 taken from Aug 2004 ($\mathrm{HJD} = 3246.5202$) to Aug 2010 ($\mathrm{HJD} = 5420.4585$) do not confirm any significant RV variations. The upper parts of Fig.~\ref{prof_hd193514} show that the Doppler shifts of the spectral lines are rather small (undetectable by eye) on several (short and long) timescales. We indeed determine $\overline{\mathrm{RVs}} = -18.0 \pm 5.2$, $-7.0 \pm 5.0$ and $-15.6 \pm 5.0$~\kms\ for the \ion{He}{i}~4471, \ion{He}{ii}~4542 and \ion{He}{i}~4713 lines. In the lower parts of Fig.~\ref{prof_hd193514}, we show the TVS computed for these different spectral lines. The variations above the dotted lines are considered as significant under a significance level of 1\%. Therefore, we see that the TVS of the \ion{He}{ii}~4542 and the \ion{He}{i}~4713 line profiles do not show any substantial variations, though the \ion{He}{i}~4471 and especially the \ion{He}{ii}~4686 lines do. However, since this variability is not detected on all the line profiles, we conclude that it is unlikely to be due to binarity but rather arises in the stellar wind itself. However, we do not have enough {\'e}chelle spectra to characterize the changes reported by \citet{und95} in the line profile of H$\alpha$. We therefore consider HD\,193514 as a presumably single star. 

From the {\'e}chelle spectra, we determine $\log W' = 0.014$ and $\log W''= 0.253$, yielding for HD\,193514 a spectral type O\,7--O\,7.5 and a luminosity class similar to a giant (III). The strong \ion{N}{iii}~4634--41 lines coupled with the \ion{He}{ii}~4686 line (for which the emission component is similar to the absorption one) suggest an (f) suffix, which agrees with a giant luminosity class. Therefore, we classify HD\,193514 as an O\,7--O\,7.5 III(f) star.

\subsubsection{HD\,193595}

HD\,193595 is located in the Berkeley\,86 young open cluster \citep{mas95}. Classified as an O\,6V \citep{gar91}, this object was poorly investigated in terms of RV.

\begin{figure}[htbp]
\begin{center}
\includegraphics[width=8.5cm,bb=18 160 575 705,clip]{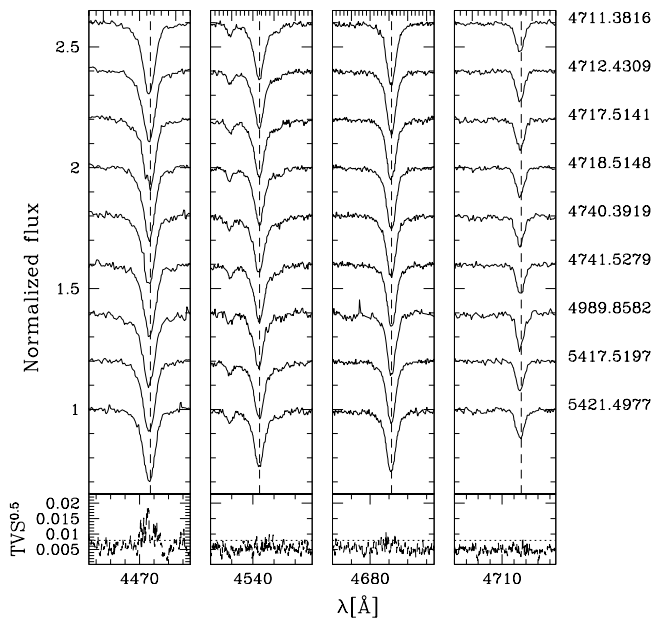}
\caption{Same as Fig.~\ref{prof_hd193514} but for HD\,193595.}\label{prof_hd193595}
\end{center}
\end{figure}

We measure the RVs on nine spectra taken at different epochs, between Sep 2008 ($\mathrm{HJD} = 4711.3816$) and Aug 2010 ($\mathrm{HJD} = 5421.4977$). These data reveal rather constant RVs for this star: $\overline{\mathrm{RVs}} = -13.7\pm2.2$, $-8.1\pm2.8$, $-4.5\pm2.6$ and $-12.5\pm1.7$~\kms\ for the \ion{He}{i}~4471, \ion{He}{ii}~4542, \ion{He}{ii}~4686 and \ion{He}{i}~4713 lines, respectively. We detect, as shown in the lower panels of Fig.~\ref{prof_hd193595}, low amplitude variations (at a significance level of 1\%) only in the TVS spectrum of the \ion{He}{i}~4471 line. The results of the TVS associated to the RV measurements thus seem to imply that the observed variations are probably not due to binarity. By assuming a single status for this object, we derive for HD\,193595 an O\,7V stellar classification ($\log W' = -0.036$ and $\log W''= 0.052$). 

\subsubsection{HD\,193682}

\begin{figure}[htbp]
\begin{center}
\includegraphics[width=8.5cm,bb=18 160 575 701,clip]{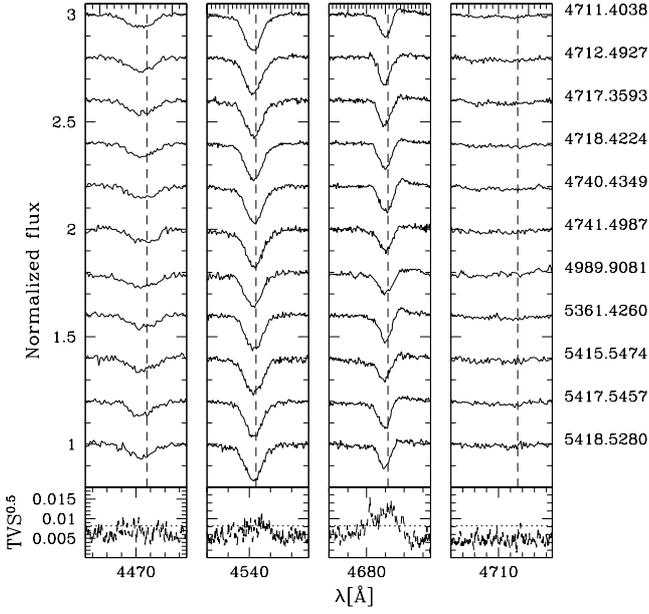}
\caption{Same as Fig.~\ref{prof_hd193514} but for HD\,193682.}\label{prof_hd193682}
\end{center}
\end{figure}

This star is the hottest object in our sample. HD\,193682 seems to have a spectral type oscillating between O\,5 \citep{hil56} and O\,4 \citep[O\,4III(f),][]{gar91}. We obtained ten spectra of this object between Sep 2008 ($\mathrm{HJD} = 4711.4038$) and Aug 2010 ($\mathrm{HJD} = 5418.5280$). These data show rather broad line profiles, indicating a rather large projected rotational velocity. We determine from the Fourier transform method of \citet{sim07} a $v \sin i = 150$~\kms\ for this object. This moderately fast rotation yields broader lines that produce larger uncertainties on their centroid and hence poorer accuracy of the RVs. However, we do not perform a least-square fit with a synthetic profile broadened by rotation to determine the RVs of this star because the shapes of the strongest lines are still Gaussian. We determine $\overline{\mathrm{RVs}}$ of about $-48.6\pm11.3$, $-39.7\pm7.7$ and $-55.1\pm11.1$~\kms\ for the \ion{He}{i}~4471, \ion{He}{ii}~4542 and \ion{He}{ii}~4686 lines, respectively. Such a high 1-$\sigma$ dispersion is probably due to the line broadening rather than to binary motion because this RV dispersion is smaller on the strongest spectral lines. Furthermore, we do not detect any secondary signature in the observed spectra nor any change of the line width as a function of time, as it is shown in the upper panels of Fig.~\ref{prof_hd193682}. Indeed, we see from this figure that the spectral lines have always their centre at the same side of the dashed lines, which represent the rest wavelengths. We compute the TVS for these spectral lines (lower panels of Fig.~\ref{prof_hd193682}) and we find a significant (although weak) variation in the line profile of \ion{He}{ii}~4686, a spectral line generally affected by strong stellar winds. We cannot consider these changes as due to binarity because they are not detected in all the spectral lines, notably in the \ion{He}{i}~4471 and the \ion{He}{ii}~4542 lines, which suggest that these variations are probably intrinsic to the stellar wind rather than attributable to a putative companion. 

We then calculate, from the \ion{He}{i}~4471--\ion{He}{ii}~4542 ratio, $\log W' = -0.514$, corresponding to an O\,5 subtype. Furthermore, Walborn's criterion suggests to add a (f) suffix, indicating a moderate emission for the \ion{N}{iii}~4634$-$41 lines and a moderate absorption for the \ion{He}{ii}~4686 line, similar to giant (III) stars. To summarize, we classify HD\,193682 as an O\,5III(f) star. This spectral classification is similar to those found in the literature.

\subsubsection{HD\,194094}

\begin{figure}[htbp]
\begin{center}
\includegraphics[width=8.5cm,bb=26 163 575 701,clip]{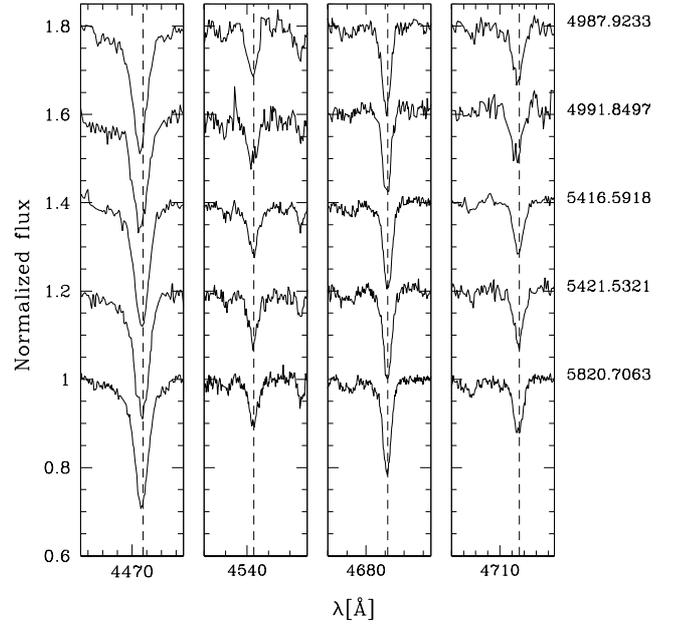}
\caption{The \ion{He}{i}~4471, \ion{He}{ii}~4542, \ion{He}{ii}~4686 and \ion{He}{i}~4713 line profiles of HD\,194094.}\label{prof_hd194094}
\end{center}
\end{figure}

During our spectroscopic campaign, HD\,194094 was only observed 5 times: twice in Jun 2009, twice in Aug 2010 and once in Sep 2011. These data however allow us to investigate RV variations on timescales from one week to one year. These five spectra do not exhibit any evidence of a putative companion on such timescales nor any periodic motion relative to the rest wavelengths (dashed lines in Fig.~\ref{prof_hd194094}). We measure the RVs on the \ion{He}{i}~4471, \ion{He}{ii}~4542, \ion{He}{ii}~4686 and \ion{He}{i}~4713 lines, reporting $\overline{\mathrm{RVs}} = -18.2\pm3.9$, $-11.3\pm3.8$, $-9.7\pm8.9$ and $-13.9\pm6.1$~\kms, respectively. The standard deviations on these measurements indicate that HD\,194094 is probably single and that the RV variations detected on the \ion{He}{ii}~4686 line are likely due to a variable stellar wind. However, unlike the stars previously analysed, the Aur{\'e}lie data for HD\,194094 are not sufficiently numerous nor sufficiently spread over the entire campaign to compute the TVS. We also cannot include the SPM data in the computation of the TVS, as we would need to convolve the data so that they all have the same resolution. Indeed, unlike HD\,46150 in \citet{mah09}, we do not have, in the [$4450-4900$]~\AA\ wavelength domain, diagnostic lines similar to the interstellar \ion{Na}{i} lines at 5890~\AA\ and 5896~\AA\ to compare the convolved spectra.

Finally, by assuming that this object is single, we compute, from an {\'e}chelle spectrum of HD\,194094, $\log W'=0.232$, $\log W'' = 0.174$ and $\log W'''= 5.204$, yielding an O\,8.5III star. This spectral type is slightly earlier than reported in the literature \citep[O\,9III,][]{hil56}. However, the error-bars on the determination of the spectral type do not allow us to claim that the new classification is significantly different from the older ones.

\subsubsection{HD\,194280}

Presented as a prototype of the OC\,9.7Iab class, HD\,194280 appears to be a carbon-rich star with a depletion in nitrogen \citep{goy73,wal00}. According to these authors, the profiles of \ion{He}{i}~5876 and H$\alpha$ would indicate that the stellar wind is weak.

\begin{figure}[htbp]
\begin{center}
\includegraphics[width=8.5cm,bb=18 160 575 705,clip]{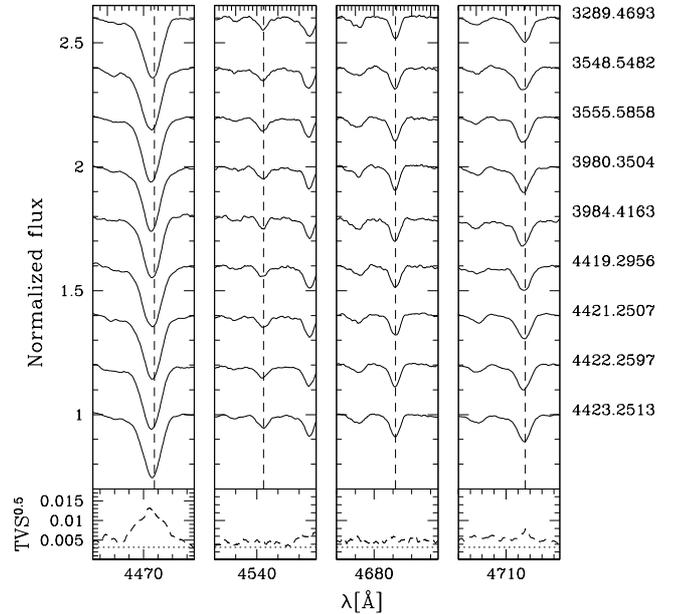}
\caption{Sample of spectra taken from the entire dataset. This figure is similar to Fig.~\ref{prof_hd193514} but for HD\,194280.}\label{prof_hd194280}
\end{center}
\end{figure}

We obtained forty-two spectra of this target with the Aur{\'e}lie spectrograph in the [$4450-4900$]~\AA\ wavelength domain between Oct 2004 ($\mathrm{HJD} = 3286.3646$) and Nov 2007 ($\mathrm{HJD} = 4423.2513$). The spectral lines sampled over different timescales are displayed in the upper panels of Fig.~\ref{prof_hd194280}. This figure shows that the Doppler shifts of the lines are small, which is confirmed by the $\overline{\mathrm{RVs}}$. These values for the \ion{He}{i}~4471, \ion{He}{ii}~4542, \ion{He}{ii}~4686 and \ion{He}{i}~4713 lines are indeed equal to $-24.9\pm4.2$, $-11.6\pm7.6$, $-5.4\pm5.0$ and $-17.1\pm4.8$~\kms, respectively, suggesting that this star is presumably single. From the \ion{He}{i}~4471--\ion{He}{ii}~4542 ratio, we obtain $\log W' = 0.742$, which agrees with the O\,9.7 subtype. Furthermore, the object presents strong line-profile variations notably in the \ion{He}{i}~4471 line. The results of the TVS are shown in the lower panels of Fig.~\ref{prof_hd194280}. Since the profile of the \ion{He}{ii}~4686 line does not appear in emission and since the profiles of the \ion{He}{i}~5876 and H$\alpha$ lines did not seem to indicate strong stellar wind \citep{goy73}, these variations are unlikely due to the wind. Therefore, another possible explanation for this strong variability in \ion{He}{i}~4471 could be pulsations.

\subsubsection{HD\,228841}

\citet{mas95} located this object in the Berkeley\,86 young open cluster, along with HD\,193595 and HD\,228989. This star was successively classified as O\,7V((f)) and O\,6.5Vn((f)) by \citet{mas95} and \citet[][and references therein]{sota11}. 

\begin{figure}[htbp]
\begin{center}
\includegraphics[width=8.5cm,bb=18 160 575 701,clip]{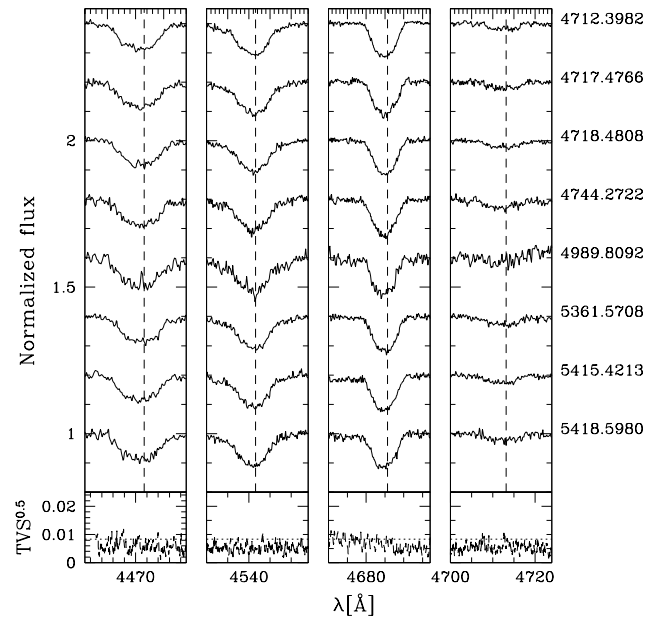}
\caption{Same as Fig.~\ref{prof_hd193514} but for HD\,228841.}\label{prof_hd228841}
\end{center}
\end{figure}

Eight optical spectra were collected from Sep 2008 ($\mathrm{HJD} =4712.3982$) to Aug 2010 ($\mathrm{HJD}=5418.5980$), mainly in the [$4450-4490$]~\AA\ range. The spectra show broad lines which means that HD\,228841 is likely a rapid rotator. We indeed see, in the upper panels of Fig.~\ref{prof_hd228841}, that the spectral line widths remain constant as a function of time, which does not seem to support the presence of a secondary component. Therefore, we determine from the Fourier transform method of \citet{sim07} the projected rotational velocity of the star. We obtain a value of $317$~\kms. The mean RVs determined by least-square fit between a rotation profile and the spectral lines amount to $\overline{\mathrm{RVs}} = -48.4\pm8.9$, $-25.5\pm6.5$ and $-44.9\pm25.3$~\kms\ for the \ion{He}{i}~4471, \ion{He}{ii}~4542 and \ion{He}{i}~4713 lines, respectively. Given the $v \sin i$, the standard deviations generally smaller than 15~\kms\ favour a single status for this object. In addition, the variations in the line profiles computed with the TVS (see the lower panels of Fig.~\ref{prof_hd228841}) are not significant, which agrees with the presumably single status for HD\,228841. The high dispersion detected for the \ion{He}{i}~4713 line is probably due to the weakness of this line.

We then derive the spectral type of this star from Conti's criterion. $\log W'$ is equal to $-0.011$, which indicates an O\,7 star, as already determined by \citet{mas95}. We stress however that $\log W''$ could not be estimated given the broadness of the lines and thus the severe blend of the \ion{Si}{iv}~4089 line with H$\delta$. 


\section{O-type stars in the Cyg\,OB3 association}
\label{sect:ob3}
\subsection{Presumably single stars}
\label{subsec:single3}

\subsubsection{HD\,190864}

Quoted as belonging to NGC\,6871 \citep{hum78}, HD\,190864 is one of the brightest O-type stars of Cyg\,OB3. The spectral classifications derived for this object range between O\,6 \citep{hil56} and O\,7 \citep{con74} but agree on a giant luminosity class. Previous investigations \citep[e.g.,][]{pla24} quoted this star as SB1 but no solution for its orbital motion has been published. \citet{mei68} listed this star as visual double star with $\Delta m=0.5$ (O\,6$+$B\,0.5Vp). Although these authors did not mention the angular separation, an B\,0.5Vp spectral type could match with the spectral type of HD\,227586 located at $1\arcmin$ from HD\,190864. More recently, \citet{gar92} attributed to HD\,190864 a spectral type of B\,1II$+$O\,6.5III(f), assigning it an SB2 status. However, it is possible that the B\,1II component is the visual component reported by \citet{mei68}. Indeed, the latter author also mentioned that the estimate of the luminosity class of the B component was uncertain and could be higher, thereby could be attributed to the B\,1II component of \citet{gar92}. From seven observations spread over two years, \citet{und95} did however not detect any significant RV variations. According to this author, it is unlikely that the difference between the projected rotational velocity given by \citet[][69~\kms]{coe77} and that of \citet[][105~\kms]{herrero1992} is due to a possible varying blend of the profiles of two putative components but rather seems to originate from an emission component in some absorption lines.

\begin{figure}[htbp]
\begin{center}
\includegraphics[width=8.5cm,bb=18 160 584 705,clip]{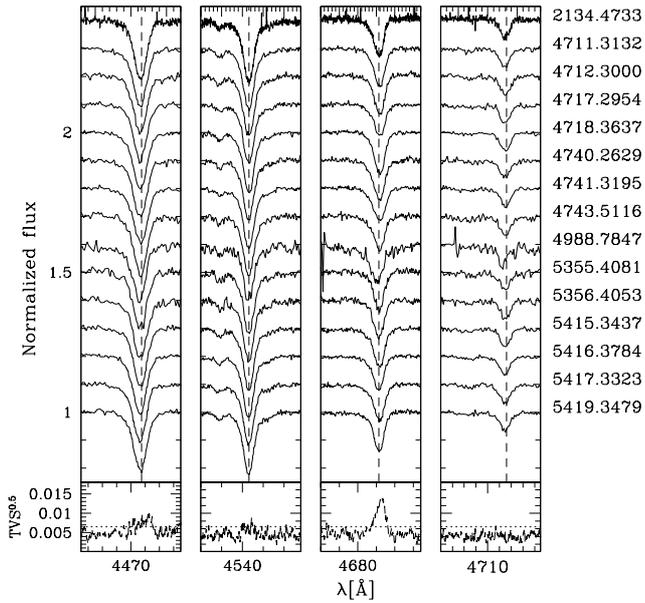}
\caption{Same as Fig.~\ref{prof_hd193514} but for HD\,190864.}\label{prof_hd190864}
\end{center}
\end{figure}

We obtained seventeen spectra between Aug 2001 ($\mathrm{HJD} = 2134.4733$) and Aug 2010 ($\mathrm{HJD} = 5419.3479$). This dataset does not show any significant RV changes, except for the \ion{He}{ii}~4686 line. The mean RV and the standard deviation measured on this line are indeed of $\overline{\mathrm{RV}} = 4.6\pm11.5$~\kms\ whilst the other values derived on the \ion{He}{i}~4471, \ion{He}{ii}~4542 and \ion{He}{i}~4713 lines amount to $\overline{\mathrm{RVs}} = -15.5 \pm 2.5$, $-6.3 \pm 2.4$ and $-16.5 \pm 3.8$~\kms, respectively. Since our observations cover both short and long timescales, we can exclude that HD\,190864 is a binary system with an orbital period shorter than about 3000 days. From an intense survey devoted to HD\,190864, we have never detected the signature of a putative companion. If a B component was present in the spectrum of HD\,190864, we should see it. According to \citet{sch82}, a B\,0.5V or a B\,1II component should be only 6 times fainter than an O\,6.5III(f) star and the mass ratio should be between 1 and 4. Therefore, such a component should be visible in the observed spectra, which is not the case. It is thus unlikely that the difference between the two values of \vsini, given by \citet{con77} and \citet{herrero1992}, originates from binarity. Furthermore, if we consider that the RV published by \citet[$-15.3$~\kms,][]{con77} was measured on a \ion{He}{i} line, this could constitute an additional clue that the star is presumably single. This thus reveals the importance of a homogeneous dataset to study the binary fraction of massive star populations (as mentioned in Sect.~\ref{sect:obs}). 

\begin{figure}[h]
\includegraphics[width=8.5cm,bb=46 425 335 682,clip]{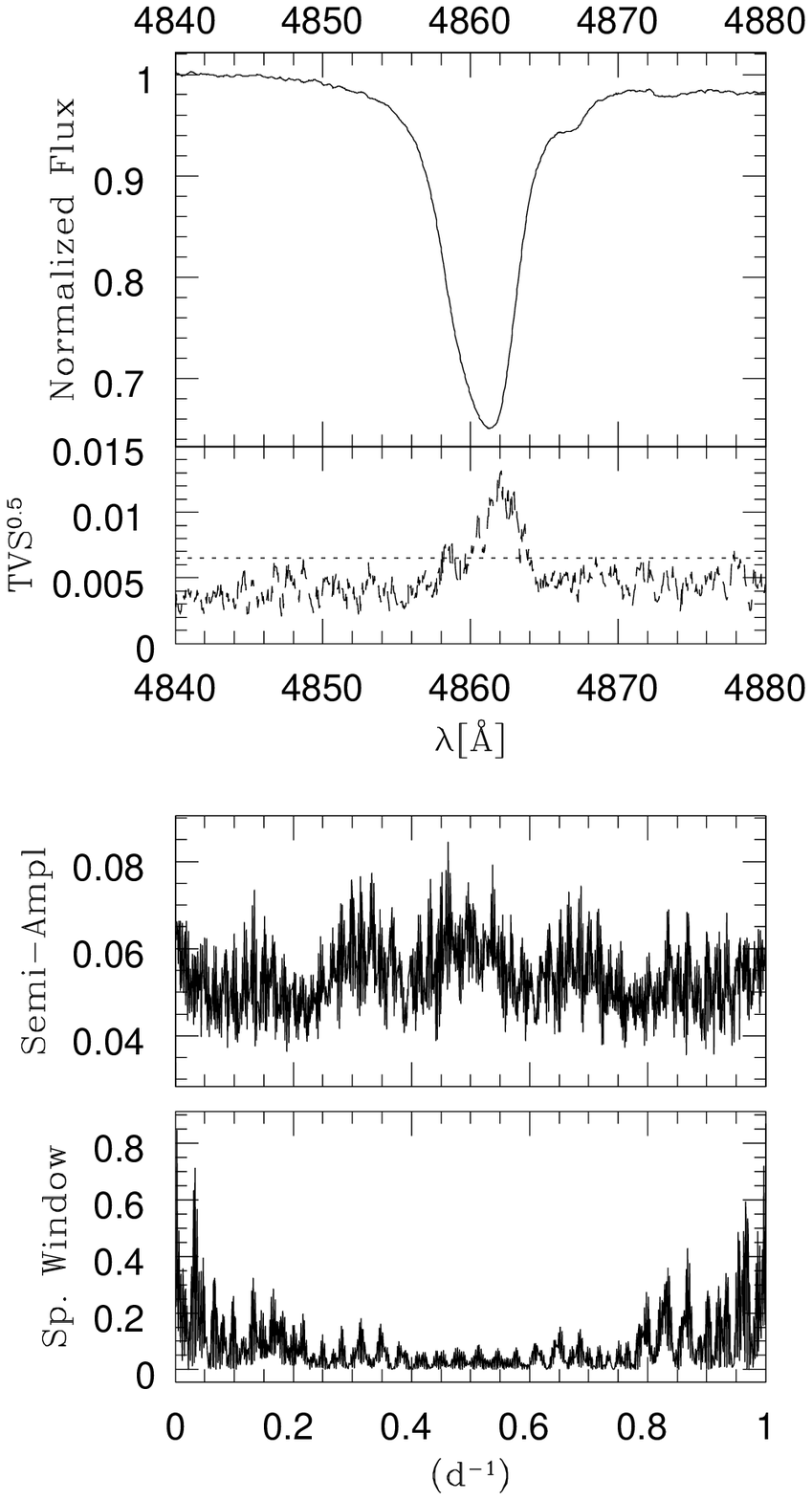}
\caption{Mean spectrum and TVS of HD\,190864 computed from the Aur{\'e}lie data of the H$\beta$ line. The dotted line illustrates the 1\% significance level for the variability evaluated following the approach of \citet{ful96}. }\label{tvs_hd190864}
\end{figure}

Though no signature of a secondary companion is found in the HD\,190864 spectrum, the TVS analysis reveals significant variations of some line profiles, mainly in the red wings of the \ion{He}{ii}~4686 and H$\beta$ lines (lower panels of Figs.~\ref{prof_hd190864}, and Fig.~\ref{tvs_hd190864}). It is however unlikely that these variations are due to the existence of a secondary star. If that was the case, the variability pattern should be once again observed in all the spectral lines which is clearly not the case. Therefore, we classify this object as presumably single.

Finally, we derive from Conti's criterion a spectral classification of O\,6.5 ($\log W' = -0.193$), consistent with the previous estimates. The presence of moderate emission in the \ion{N}{iii}~4634--41 lines and a rather weak \ion{He}{ii}~4686 line suggest to add the (f) suffix. Therefore, we derive that the spectral classification for HD\,190864 is O\,6.5III(f). 

\subsubsection{HD\,227018}

Located at about $1.2\degr$ south-west of HD\,190864, HD\,227018 is rather faint ($V=8.98$) for our survey. While its spectral type is relatively well known (O\,6.5--O\,7), its luminosity class is poorly constrained. Indeed, \citet{herrero1992} quoted this star as an O\,6.5III while \citet{mas95} attributed it an O\,7V((f)) type.  

\begin{figure}[htbp]
\begin{center}
\includegraphics[width=8.5cm,bb=18 160 575 705,clip]{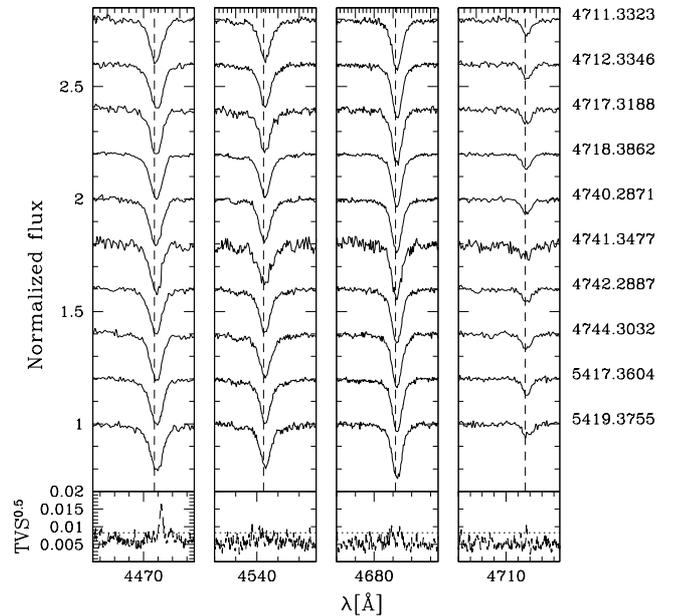}
\caption{Same as Fig.~\ref{prof_hd193514} but for HD\,227018.}\label{prof_hd227018}
\end{center}
\end{figure}

Between Sep and Oct 2008, we collected eight spectra with the Aur{\'e}lie spectrograph and we increased our dataset with two additional spectra taken in August 2010 with the same instrument. The measurements of the RVs reveal rather constant values with $\overline{\mathrm{RVs}} = 15.6 \pm 3.4$, $21.7 \pm 4.0$, $23.8 \pm 3.3$ and $16.6 \pm 3.9$~\kms\ for the \ion{He}{i}~4471, \ion{He}{ii}~4542, \ion{He}{ii}~4686 and \ion{He}{i}~4713 lines, respectively. We see from the upper panels of Fig.~\ref{prof_hd227018} that the Doppler shifts of the spectral lines are relatively small relative to the rest wavelengths. Moreover, the TVS spectra (lower panels of Fig.~\ref{prof_hd227018}) only exhibit small variations in the \ion{He}{i}~4471 line at the limit of being significant while the other lines are relatively stable, thereby indicating that HD\,227018 is presumably single.
In order to derive the spectral type of this star, we compute, from the \ion{He}{i}--\ion{He}{ii} ratio, $\log W' = -0.201$ which corresponds to an O\,6.5 star. We add the ((f)) suffix reminiscent of a main-sequence star because the object shows weak \ion{N}{iii}~4634--41 emissions and strong \ion{He}{ii}~4686 absorption lines. Therefore, we classify this star as O\,6.5V((f)).

\subsubsection{HD\,227245}

\begin{figure}[htbp]
\begin{center}
\includegraphics[width=8.5cm,bb=26 165 575 701,clip]{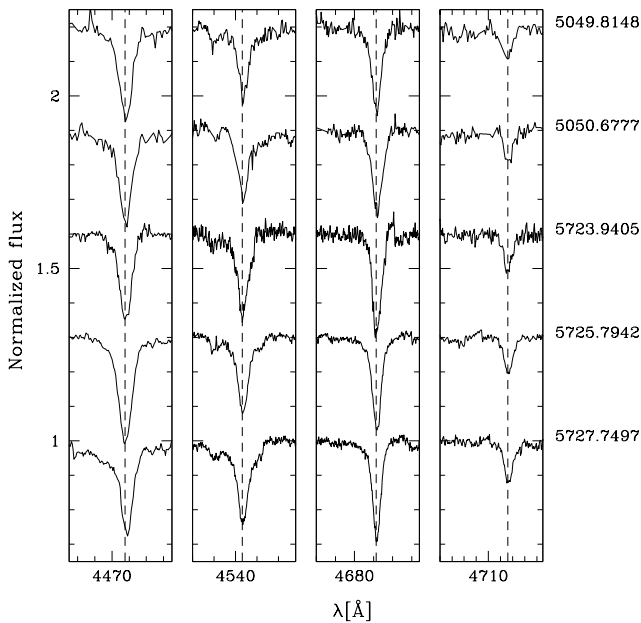}
\caption{The \ion{He}{i}~4471, \ion{He}{ii}~4542, \ion{He}{ii}~4686 and \ion{He}{i}~4713 line profiles of HD\,227245.}\label{prof_hd227245}
\end{center}
\end{figure}

HD\,227245 is known as an O\,7V \citep{gar91} but its multiplicity has never been investigated until now. From five observations taken between Aug 2009 ($\mathrm{HJD} = 5049.8148$) and Jun 2011 ($\mathrm{HJD} = 5727.7497$), we determine $\overline{\mathrm{RVs}} = 5.9 \pm 5.4$, $9.8 \pm 3.9$, $10.8 \pm 6.2$ and $6.0 \pm 7.3$ for the \ion{He}{i}~4471, \ion{He}{ii}~4542, \ion{He}{ii}~4686 and \ion{He}{i}~4713 lines, respectively. Figure~\ref{prof_hd227245} shows these spectral lines at different HJD. We stress that a normalization problem occurs at HJD~=~5723.9405 for the \ion{He}{i}~4471 line but it does not affect the determination of the RV for this line. The standard deviations computed from the RVs are not considered significant because they are generally smaller (except for \ion{He}{i}~4713) than the $7-8$~\kms\ threshold defined as our variability criterion. The exception for the \ion{He}{i}~4713 line can be explained by the weakness of this line and the signal-to-noise ratio of the data. HD\,227245 is thus considered as presumably single.

The spectrum of HD\,227245 exhibits \ion{He}{ii} slightly weaker than \ion{He}{i}, indicating a mid O-type star. We obtain $\log W'= -0.034$ and $\log W'' = 0.218 $, corresponding to a spectral type O\,7III. However, the weak \ion{N}{iii}~4634--41 lines coupled with a strong \ion{He}{ii}~4686 line suggest an ((f)) suffix, indicating rather a main-sequence luminosity class. This difference could come either from the definition of the continuum level in the normalized spectra or because Conti's criterion is barely applicable to this spectral type. Therefore, we classify HD\,227245 as an O\,7V--III((f)) star. This spectral classification is in agreement with that given by \citet{gar91}.

\subsubsection{HD\,227757}

\begin{figure}[htbp]
\begin{center}
\includegraphics[width=8.5cm,bb=26 165 575 701,clip]{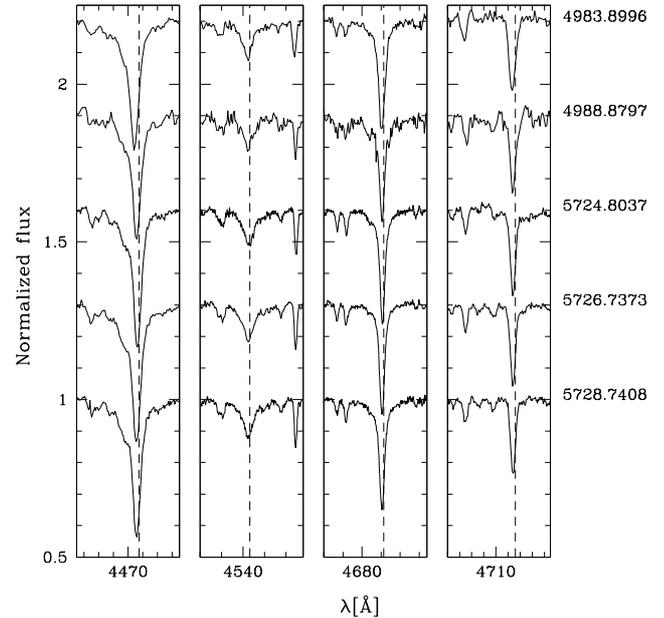}
\caption{The \ion{He}{i}~4471, \ion{He}{ii}~4542, \ion{He}{ii}~4686 and \ion{He}{i}~4713 line profiles of HD\,227757.}\label{prof_hd227757}
\end{center}
\end{figure}

HD\,227757 ($V = 9.25$) was reported by \citet{gar92} and \citet{herrero1992} to be an O\,9.5V star. From six spectra taken between Jun 2009 ($\mathrm{HJD}=4988.8797$) and Jun 2011 ($\mathrm{HJD} = 5728.7408$), we do not observe significant variations in the RVs of HD\,227757 on short nor on long timescales (see Fig.~\ref{prof_hd227757}). Indeed, we measure $\overline{\mathrm{RVs}} = -26.8\pm4.6$, $-26.2\pm4.5$, $-28.7\pm2.9$ and $-30.0\pm4.3$~\kms\ for the \ion{He}{i}~4471, \ion{He}{ii}~4542, \ion{He}{ii}~4686 and \ion{He}{i}~4713 lines, respectively. We thus consider the star as presumably single. 

The spectrum of HD\,227757 displays a stronger \ion{He}{i}~4471 line than the \ion{He}{ii}~4542 line, indicating a spectral type \citep{Walborn1990} later than O\,7. Due to the late subtype and the small \vsini\ of the star ($\sim 45$~\kms), we can clearly distinguish an \ion{O}{ii} line blended with the \ion{He}{i}~4471 line. By removing the contribution of the former line (i.e., by fitting two Gaussian profiles, one on the \ion{O}{ii} line and one on the \ion{He}{i} line), we measure $\log W' = 0.332$, corresponding to an O\,9 star. We also obtain $\log W''= 0.081$ and $\log W'''= 5.485$ which indicate a main-sequence luminosity class (V). Therefore, we classify HD\,227757 as O\,9V.


\section{O-type stars in the Cyg\,OB8 association}
\label{sect:ob8}
\subsection{Presumably single stars}
\label{subsec:single8}

\subsubsection{HD\,191423}

Better known under the name of Howarth's star, HD\,191423 is the fastest rotator known to date among Galactic O stars and it was quoted as the prototype of the ONn stars \citep[nitrogen-rich O star with broad diffuse lines,][]{wal03}. Its projected rotational velocity was estimated to be around $400$~\kms\ (e.g., \citealt{pen09} computed a \vsini\ in the range of [$336-436$]~\kms). 

\begin{figure}[htbp]
\begin{center}
\includegraphics[width=8.5cm,bb=18 160 575 701,clip]{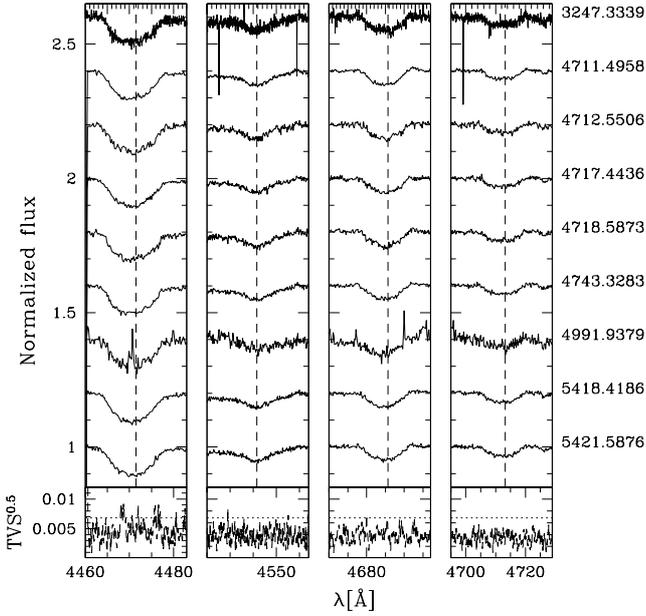}
\caption{Same as Fig.~\ref{prof_hd193514} but for HD\,191423.}\label{prof_hd191423}
\end{center}
\end{figure}

We obtained nine spectra of this star between Aug 2004 ($\mathrm{HJD}=3247.3339$) and Aug 2010 ($\mathrm{HJD}=5421.5876$). From the Fourier transform method of \citet{sim07}, we determine $v \sin i = 410$~\kms, in good agreement with previous estimates. Moreover, we see in the time series of the spectra (shown in upper panels of Fig.~\ref{prof_hd191423}) that the line widths remain constant as a function of time, thereby suggesting rather a rapid rotator than a binary system. The RVs measured by least-square fit give us average values of $\overline{\mathrm{RVs}} = -44.0 \pm 3.0$, $-58.8 \pm 18.4$, $-48.7 \pm 5.0$ and $-56.5 \pm 4.8$~\kms\ for the \ion{He}{i}~4471, \ion{He}{ii}~4542, \ion{He}{ii}~4686 and \ion{He}{i}~4713 lines, respectively. The RV uncertainties for rapid rotators are typically larger because the centroids of the lines are not well defined. This could explain the large standard deviation observed for the weak \ion{He}{ii}~4542 line whilst the other lines are rather stable in RV, thereby indicating a presumably single star. Moreover, after having computed the TVS spectra for these characteristic lines only from the Aur{\'e}lie data (lower panels of Fig.~\ref{prof_hd191423}), we detect no significant variation of line profiles, which agrees with the likely single nature of HD\,191423. 

Therefore, we compute $\log W' = 0.396$ and $\log W''' = 5.187$, assigning to this star an O\,9III type. We are not able to estimate $\log W''$ given the blend of the \ion{Si}{iv}~4089 line with H$\delta$. We also add the suffixes $n$ and N, because of the broadness of the lines as well as the strong relative intensity of the nitrogen lines, thereby giving an ON\,9IIIn spectral classification for HD\,191423. 

\subsubsection{HD\,191978}

HD\,191978 was classified as an O\,8 star by \citet{goy73}. The investigation of its RVs by \citet{abt72} did not reveal any significant variations. 

\begin{figure}[htbp]
\begin{center}
\includegraphics[width=8.5cm,bb=18 160 575 701,clip]{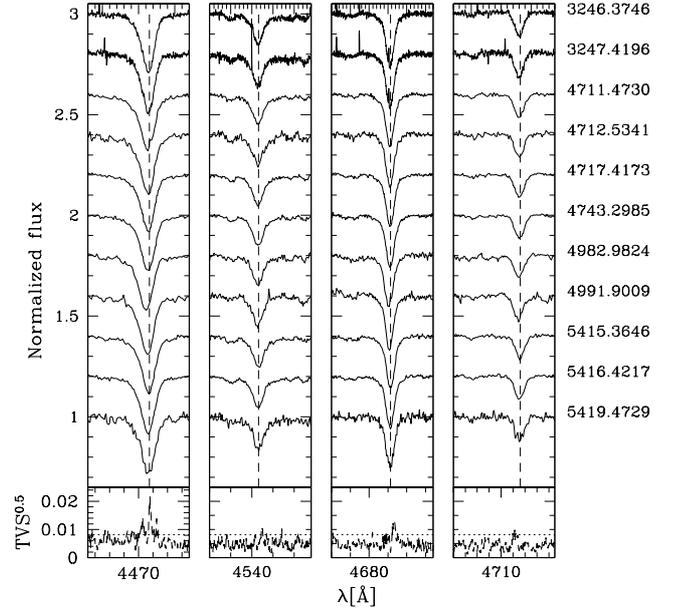}
\caption{Same as Fig.~\ref{prof_hd193514} but for HD\,191978.}\label{prof_hd191978}
\end{center}
\end{figure}

We collected eleven spectra of HD\,191978 over a timescale of about 2170 days between Aug 2004 ($\mathrm{HJD} = 3247.4196$) and Aug 2010 ($\mathrm{HJD}=5419.4729$). The measurements made on these data show relatively constant RVs. We can see in the upper panels of Fig.~\ref{prof_hd191978} that no clear Doppler shift exists relative to the rest wavelengths of four spectral lines. The mean RVs and their standard deviations are $\overline{\mathrm{RV}} = -18.1 \pm 3.9$~\kms\ for the \ion{He}{i}~4471 line, $\overline{\mathrm{RV}} = -10.4 \pm 5.3$~\kms\ for the \ion{He}{ii}~4542 line, $\overline{\mathrm{RV}} = -9.5 \pm 5.0$~\kms\ for the \ion{He}{ii}~4686 line, and $\overline{\mathrm{RV}} = -11.8 \pm 6.2$~\kms\ for the \ion{He}{i}~4713 line. However, the TVS spectra (lower panels of Fig.~\ref{prof_hd191978}) indicate significant line profile variations for the \ion{He}{i}~4471 and marginal ones for the \ion{He}{ii}~4686 line, whilst insignificant line profile variations are found for the \ion{He}{ii}~4542 and \ion{He}{i}~4713 lines. Since these variations are not detected in all the line profiles, it is not likely that they are linked to binarity. Therefore, we consider this star as presumably single. The determinations of the EWs for the diagnostic lines yield $\log W' = 0.120$, $\log W'' = 0.227$ and $\log W''' = 5.398$, which correspond to an O\,8III star. 

\subsubsection{HD\,193117}

\begin{figure}[htbp]
\begin{center}
\includegraphics[width=8.5cm,bb=18 160 575 701,clip]{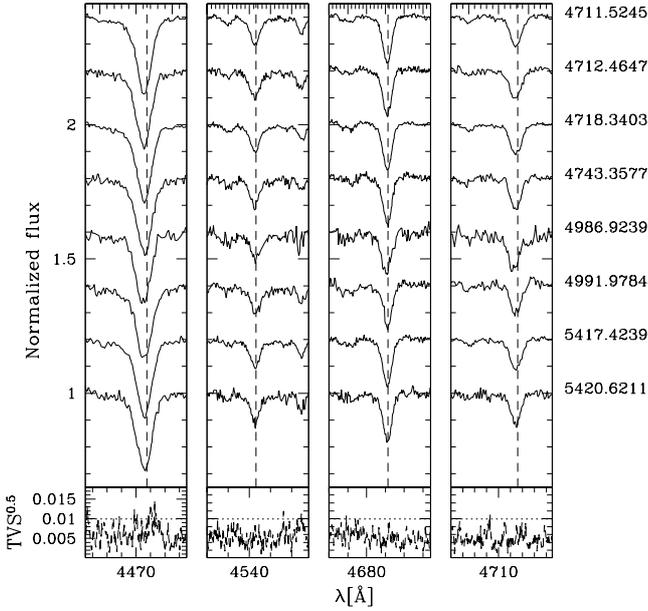}
\caption{Same as Fig.~\ref{prof_hd193514} but for HD\,193117.}\label{prof_hd193117}
\end{center}
\end{figure}

As several O stars in our sample, HD\,193117 has received little attention in the past years. We obtained eight spectra between Sep 2008 ($\mathrm{HJD} = 4711.5245$) and Aug 2010 ($\mathrm{HJD}=5420.6211$). Quoted as an O\,9.5II by \citet{goy73}, this star does not present any significant variations in its RVs. This can be seen in the upper panel of Fig.~\ref{prof_hd193117}. We indeed measure $\overline{\mathrm{RVs}} = -29.1\pm2.4$, $-16.6\pm7.1$, $-13.5\pm4.8$ and $-26.0\pm4.7$~\kms\ for the \ion{He}{i}~4471, \ion{He}{ii}~4542, \ion{He}{ii}~4686 and \ion{He}{i}~4713 lines, respectively. In addition to these results, the TVS (lower panels of Fig.~\ref{prof_hd193117}) shows no significant change in the line profiles of the star. Since we find no direct evidence of variations in the spectrum of HD\,193117, we report this star as presumably single. 

The diagnostic line ratios, \ion{He}{i}~4471--\ion{He}{ii}~4542 and \ion{Si}{iv}~4089--\ion{He}{i}~4143, give $\log W' = 0.343$ and $\log W''= 0.256$. These results correspond to an O\,9III star. Furthermore, we computed $\log W'''= 5.155$ which agrees with this spectral classification. 


\section{O-type stars in the Cyg\,OB9 association}
\label{sect:ob9}
\subsection{Gravitationally bound systems}
\label{subsec:bound9}
\subsubsection{HD\,194649}

\begin{figure}[htbp]
\begin{center}
\includegraphics[width=7cm,bb=32 164 579 704,clip]{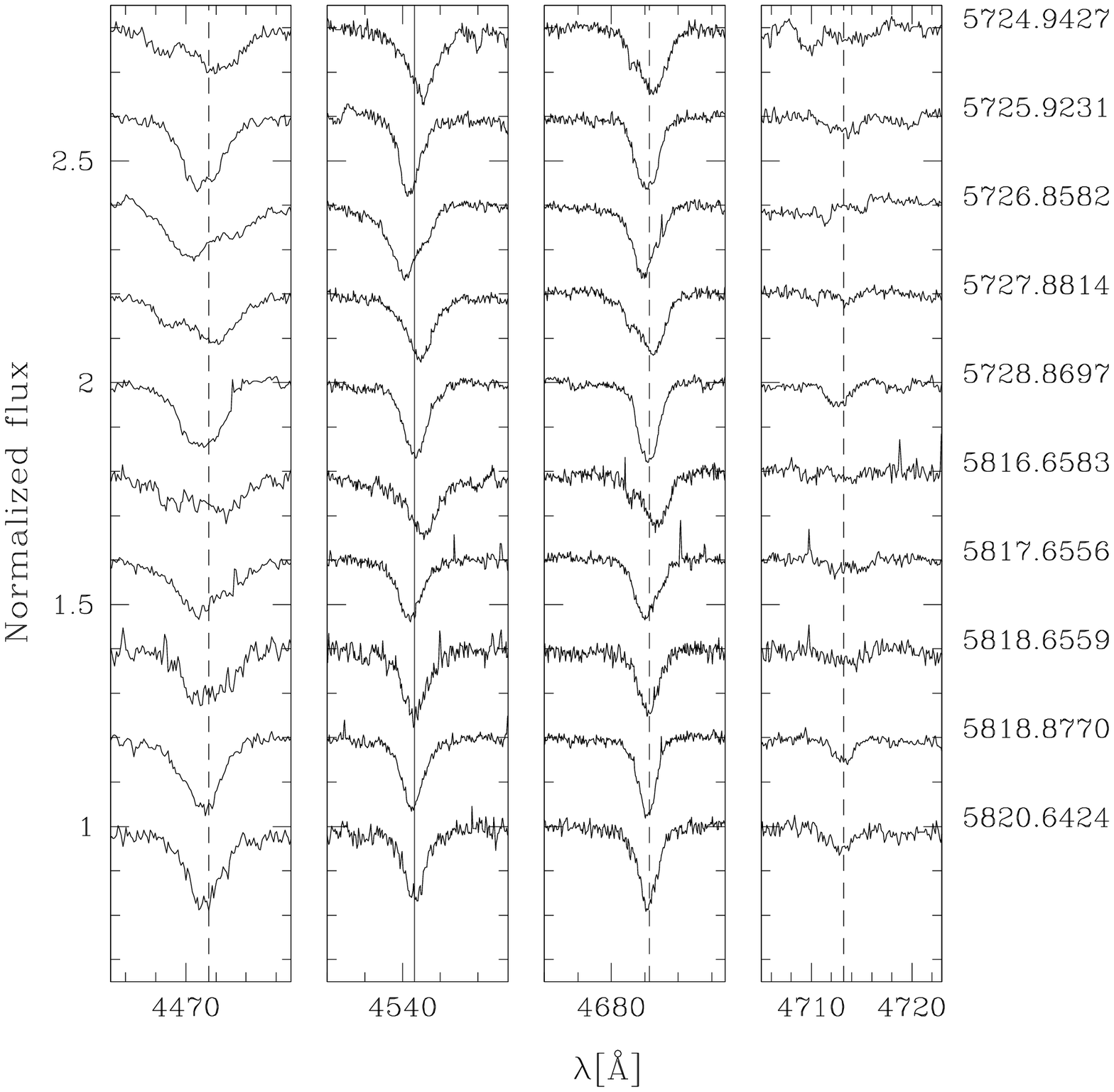}
\includegraphics[width=7cm,bb=32 164 555 676,clip]{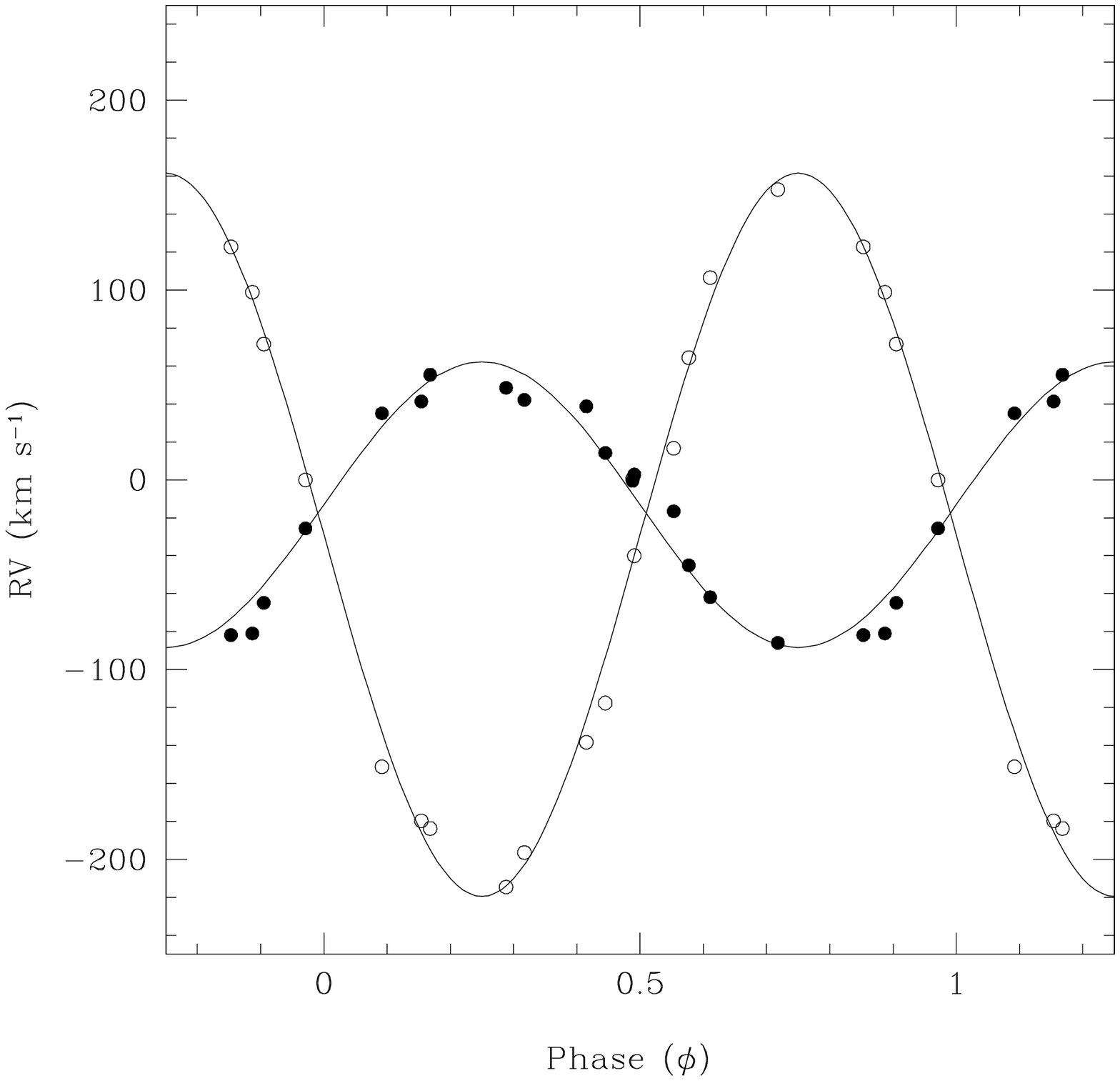}
\caption{{\it Top:} \ion{He}{i}~4471, \ion{He}{ii}~4542, \ion{He}{ii}~4686, \ion{He}{i}~4713 line profiles at different epochs. {\it Bottom:} RV curves of HD\,194649 computed from $P_{\mathrm{orb}}=3.39294$~days. Filled circles represent the primary whilst the open circles correspond to the secondary. }\label{sol_hd194649}
\end{center}
\end{figure}

Often quoted as an O\,6.5 star, HD\,194649 was reported by \citet{mul54} to be a binary system. However, no orbital solution is found in the literature. Our dataset is composed of seventeen spectra taken between Jun 2009 ($\mathrm{HJD} = 4992.8930$) and Sep 2011 ($\mathrm{HJD} = 5820.6424$). These data reveal a clear SB2 signature with a secondary component moving in anti-phase relative to the primary. This motion is clearly observed over two consecutive nights, indicating a short-period binary system. By applying the Fourier method of HMM to the RVs refined by the disentangling programme, we derive an orbital period of $3.39294 \pm 0.00139$~days. As we did for the previous binary systems, we compute two orbital solutions: an eccentric one and a circular one. However, the best orbital solution (i.e., with the smallest rms) is achieved for a circular orbit. The orbital parameters from this solution are given in Table~\ref{sol_orb} and the RV curves are shown in Fig.~\ref{sol_hd194649}. The primary appears to be significantly more massive than the secondary. Moreover, we have a difference of about 16~\kms\ between the systemic velocities of both components. This discrepancy could indicate that the secondary lines are more affected by the velocity field of a strong stellar wind, although such an interpretation is at odds with the spectral types derived below which suggest an earlier and more evolved primary star.

\begin{figure}[htbp]
\begin{center}
\includegraphics[width=8.5cm,bb=20 420 569 695,clip]{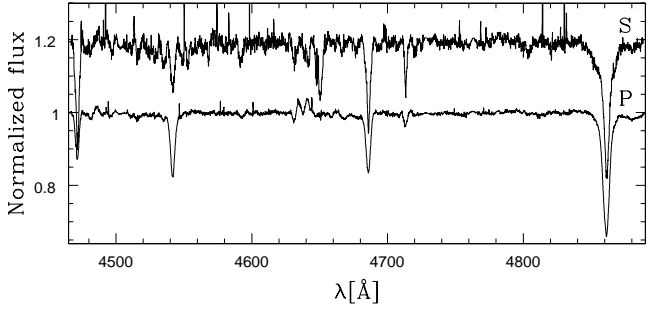}
\caption{Disentangled spectra of the two components of HD\,194649. The spectra are normalized accounting for the brightness ratio of 4.7. The secondary spectrum is vertically shifted for clarity.}\label{cmfgen_hd194649}
\end{center}
\end{figure}

We measure the EWs to determine the spectral classification on the disentangled and observed spectra. We find $\log W' = -0.267$ and $\log W' = 0.124$ for the primary and secondary components, respectively, corresponding to O\,6 and O\,8 subtypes. For the primary, it is not possible to compute $\log W''$ because the star is too early for Conti's criterion. Therefore, by focusing on the disentangled spectra (not yet corrected for the brightness ratio), the rather moderate \ion{N}{iii}~4634--41 and \ion{He}{ii}~4686 lines suggest to add an (f) tag to the primary spectral type, thus suggesting a giant luminosity class. For the secondary, we determine $\log W'' = -0.03$, corresponding to a main-sequence star (V). We thus conclude that the spectral classifications for both stars are O\,6III(f) and O\,8V for the primary and the secondary, respectively. On the basis of these spectral classifications, we derive a brightness ratio of about $4.7\pm0.4$, which agrees with a primary star more evolved than its companion. Finally, we use this brightness ratio to correct the individual spectra obtained by disentangling. The resulting spectra are displayed in Fig.~\ref{cmfgen_hd194649}.

From such spectral classifications, the tables of \citet{mar05} indicate radii and masses of $15$~\rsun\ and $35$~\msun\ for an O\,6III(f) star and of $8$~\rsun\ and $21$~\msun\ for an O\,8V star. From these masses and the minimum values provided in Table~\ref{sol_orb}, we compute an inclination between $26^{\circ}$ and $32^{\circ}$ for the system. Furthermore, we determine, on the basis of \citet{egg83}, $RRL \sin i$ of $8.3$~\rsun\ and $5.4$~\rsun\ for the primary and the secondary, respectively, thereby giving a radius between $16$~\rsun\ and $19$~\rsun\ for the primary Roche lobe and between $10$~\rsun\ and $12$~\rsun\ for the secondary one. If the standard values are close to those of both stars, it thus appears unlikely that both components of HD\,194649 fill their Roche lobe.

\subsection{Presumably single stars}
\label{subsec:single9}
\subsubsection{HD\,194334}

\begin{figure}[htbp]
\begin{center}
\includegraphics[width=8.5cm,bb=18 160 575 701,clip]{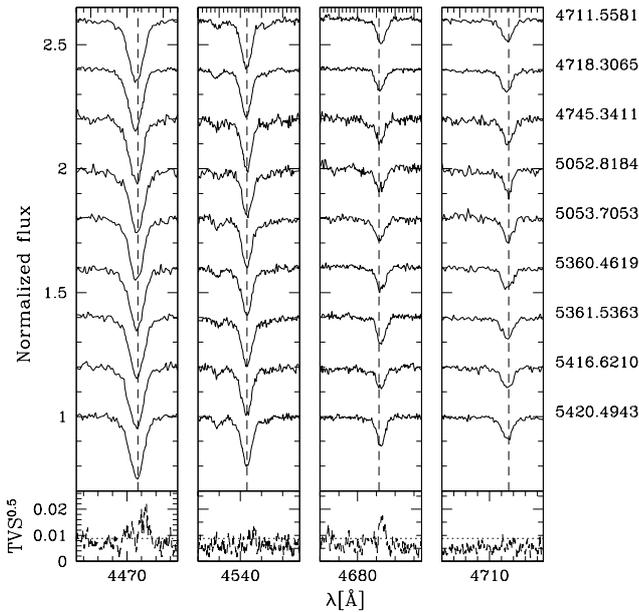}
\caption{Same as Fig.~\ref{prof_hd193514} but for HD\,194334.}\label{prof_hd194334}
\end{center}
\end{figure}

HD\,194334 was classified as an O\,7.5V star by \citet{goy73}. We collected eleven spectra between Sep 2008 ($\mathrm{HJD}=4711.5581$) and Aug 2010 ($\mathrm{HJD}=~5420.4943$). We compute $\overline{\mathrm{RVs}} = -15.3 \pm 4.6$, $-3.8 \pm 5.2$, $23.1 \pm 12.7$ and $-15.6 \pm 5.3$~\kms, for the \ion{He}{i}~4471, \ion{He}{ii}~4542, \ion{He}{ii}~4686 and \ion{He}{i}~4713 lines, respectively. These velocities reveal a significant variability in the \ion{He}{ii}~4686 line while the other lines are rather constant (upper panels of Fig.~\ref{prof_hd194334}). The mean RV of the \ion{He}{ii}~4686 line is positive whilst the others are negative. This could suggest that this line is formed in the wind. Moreover, changes in certain line profiles are also detected in the TVS spectra (lower panels of Fig.~\ref{prof_hd194334}), as notably for the \ion{He}{i}~4471 and \ion{He}{ii}~4686 lines. However, no similar change is observed in the \ion{He}{ii}~4542 and \ion{He}{i}~4713 lines, which strengthens the assumption that these variations are likely produced in the stellar wind. By assuming that HD\,194334 is a single star, we determine $\log W' = -0.002$ and $\log W''=0.294$, thus giving for HD\,194334 an O\,7--7.5III spectral classification. The luminosity class is confirmed by Walborn's criterion. Indeed, the spectrum of HD\,194334 shows moderate \ion{N}{iii}~4634--41 emissions and \ion{He}{ii}~4686 absorption lines, thereby suggesting a giant luminosity class. Therefore, HD\,194334 is classified as an O\,7-O\,7.5III(f) star.

\subsubsection{HD\,195213}

\begin{figure}[htbp]
\begin{center}
\includegraphics[width=8.5cm,bb=18 160 575 701,clip]{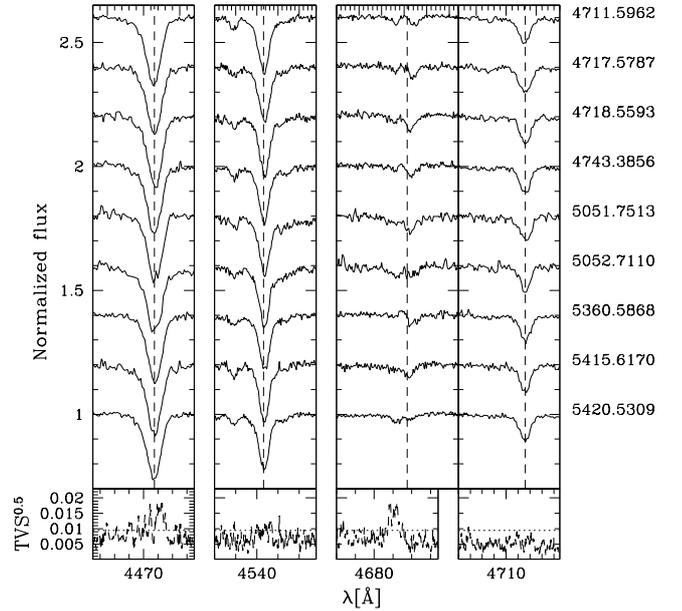}
\caption{Same as Fig.~\ref{prof_hd193514} but for HD\,195213.}\label{prof_hd195213}
\end{center}
\end{figure}

Reported as an O\,7 star by \citet{goy73}, HD\,195213 has been observed ten times during our campaign between Sep 2008 ($\mathrm{HJD} = 4711.5962$) and Aug 2010 ($\mathrm{HJD} = 5420.5309$). Our data do not reveal any significant variations of the Doppler shifts for the main lines (see upper panels of Fig.~\ref{prof_hd195213}). We indeed compute $\overline{\mathrm{RVs}} = 2.1\pm4.5$, $9.6\pm4.6$ and $-2.0\pm4.2$~\kms\ for the \ion{He}{i}~4471, \ion{He}{ii}~4542 and \ion{He}{i}~4713 lines, respectively. The \ion{He}{ii}~4686 line exhibits a pattern with bf a strong emission in the line core that almost completely masks the absorption profile, thereby probably indicating that the stellar wind of the star is relatively strong. We detect, through the TVS analysis (lower panels of Fig.~\ref{prof_hd195213}), significant variations for the \ion{He}{i}~4471 and \ion{He}{ii}~4686 lines. Since the TVS does not indicate line profile variations in all the lines and since the RV variations are not significant, we assume that the observed variability is likely due to the stellar wind rather than to a putative companion.

From the {\'e}chelle spectrum, we measure $\log W' = -0.026$, corresponding to an O\,7 star and $\log W'' = 0.264$, indicating a giant luminosity class (III). Moreover, we add the (f) suffix to the spectral type because the \ion{N}{iii}~4634--41 emissions are strong whilst the \ion{He}{ii}~4686 absorption is rather weak. We thus suggest an O\,7III(f) spectral classification for this star. 

\setcounter{table}{2}
\begin{table*}
\begin{center}
\caption{Orbital solutions for the four newly-detected binary systems \label{sol_orb}}
\begin{tabular}{|l|rr|rr|c|rr|}
\hline\hline
Parameters &  \multicolumn{2}{|c|}{HD\,193443} & \multicolumn{2}{|c|}{HD\,228989} & HD\,229234 & \multicolumn{2}{|c|}{HD194649}\\
           &  Primary            & Secondary & Primary      &     Secondary   & Primary    & Primary & Secondary\\
\hline
$P$ [d] & \multicolumn{2}{|c|}{7.467 $\pm$ 0.003}                     &  \multicolumn{2}{|c|}{1.77352 $\pm$ 0.00041} & 3.51059 $\pm$ 0.00175    & \multicolumn{2}{|c|}{3.39294 $\pm$ 0.00139}\\
$e$& \multicolumn{2}{|c|}{$0.315 \pm 0.024$}                          &   \multicolumn{2}{|c|}{$0.0$}        & $0.0$            & \multicolumn{2}{|c|}{$0.0$}\\
$\omega~[^{\circ}]$& \multicolumn{2}{|c|}{$289.7 \pm 5.8$}             &          &           & \\
$T_0$ (HJD)& \multicolumn{2}{|c|}{3246.120 $\pm$ 0.106}               & \multicolumn{2}{|c|}{5047.771 $\pm$ 0.003}   & 4714.326 $\pm$ 0.011     & \multicolumn{2}{|c|}{4991.082 $\pm$ 0.009}\\
$\gamma$ [\kms]& -10.0 $\pm$ 1.2 & -16.6 $\pm$ 2.0                  & $-3.7 \pm 1.8$ & $-7.8 \pm 1.9$            & -16.3 $\pm$ 0.8          & -13.8 $\pm$ 2.1 & -29.5 $\pm$ 3.1\\          
$K$ [\kms]& 38.8 $\pm$ 1.5 & 91.3 $\pm$ 3.4                         &  194.3 $\pm$ 2.5 & 220.3 $\pm$ 2.9         & 48.5 $\pm$ 1.1           & 76.3 $\pm$ 2.5 & 192.1 $\pm$ 6.2\\ 
$a\,\sin\,i$ [$\rsun$] & 5.4 $\pm$ 0.2 & 12.8 $\pm$ 0.5             & 6.8 $\pm$ 0.1 & 7.7 $\pm$ 0.1              & 3.4 $\pm$ 0.1            & 5.1 $\pm$ 0.2 & 12.9 $\pm$ 0.4\\
$M\,\sin^3\,i$ [$\msun$] & 1.0 $\pm$ 0.1 & 0.4 $\pm$ 0.1            & 7.0 $\pm$ 0.2 & 6.1 $\pm$ 0.2              &                          & 4.9 $\pm$ 0.4 & 1.9 $\pm$ 0.1\\ 
$q$ ($M_{\mathrm{P}}/M_{\mathrm{S}}$)& \multicolumn{2}{|c|}{2.35 $\pm$ 0.11} & \multicolumn{2}{|c|}{1.13 $\pm$ 0.02}       &                          & \multicolumn{2}{|c|}{2.52 $\pm$ 0.13} \\
$f_{\mathrm{mass}}$ [$\msun$]      &                 &                    &                  &                          &0.042 $\pm$ 0.003          &             &                  \\     
rms [\kms] & \multicolumn{2}{|c|}{5.08}                               & \multicolumn{2}{|c|}{8.37}                   & 3.04                     & \multicolumn{2}{|c|}{11.13}\\
\hline
\end{tabular}
\end{center}
\tablefoot{The 1-$\sigma$ error-bars on the parameters are provided by LOSP, except for the orbital period whose uncertainty is determined on the basis of the natural width of the peaks in the Fourier power spectrum. The orbital solution of HD\,193443 was computed assuming that the uncertainties on the secondary RVs are three times as large as for the primary.}
\end{table*}


\section{Discussion}
\label{sect:disc}
\subsection{Observational biases}

We performed Monte-Carlo simulations to estimate the probability to detect binary systems on the basis of our temporal sampling, for the 15 presumably single stars. For that purpose, we randomly draw the orbital parameters of 100000 binary systems. We consider that these systems are not detected if the standard deviation of the RVs is smaller than $7-8$~\kms\ to be consistent with our binarity criterion. For the fast rotators, we request standard deviations of at least $15$~\kms. The period distribution is chosen to be bi-uniform in log scale as in \citet{san09}. Under this assumption, $\log P$ is following a bi-uniform distribution between $0.3$ and $1.0$ for 60\% of the systems and between $1.0$ and $3.5$ for the remaining 40\%. The eccentricities are selected uniformly between 0.0 and 0.9. As \citet{rau11} already did, we assume that, for orbital periods shorter than 4 days, the systems have circular orbits. Indeed, with such periods, the circularization of systems composed of O-type stars is generally already done or almost achieved. The mass ratio ($M_P/M_S$) is uniformly distributed between 1.0 and 10.0 and the mass of the primary is obtained from tables of \citet{mar05} according to the spectral classification of the star. The longitudes of periastron are uniformly drawn between 0 and $2\pi$, and the orbital inclinations are randomly drawn according to $ \cos i \in [-1;1]$ from a uniform distribution.

\begin{table}[htbp]
\footnotesize
\begin{center}
\caption{Binary detection probability (\%) for the time sampling associated to the different presumably single stars in our sample and for various period ranges (expressed in days).}\label{cygnus_bias}
\begin{tabular}{lcccc}
\hline\hline
Stars	&   Short    & Intermed.    & Long              &  All \\	    
        &   [$2-10$] & [$10-365$]   & [$365-3165$]	&  [$2-3165$]  \\
\hline
HD\,193514 & 99.8 & 96.0 & 84.1 & 94.1 \\
HD\,193595 & 99.7 & 86.6 & 64.5 & 86.6 \\
HD\,193682 & 99.8 & 92.4 & 73.6 & 90.2 \\
HD\,194094 & 99.2 & 87.0 & 68.8 & 87.8 \\
HD\,194280 & 99.9 & 96.0 & 85.1 & 94.5 \\
HD\,228841 & 99.0 & 81.4 & 44.3 & 78.8 \\
HD\,190864 & 99.9 & 94.5 & 76.0 & 91.2 \\
HD\,227018 & 99.8 & 82.9 & 57.1 & 84.0 \\
HD\,227245 & 99.5 & 68.3 & 59.5 & 83.9 \\
HD\,227757 & 99.5 & 82.4 & 63.3 & 86.0 \\
HD\,191423 & 98.8 & 72.3 & 36.5 & 75.6 \\
HD\,191978 & 99.8 & 94.1 & 79.5 & 92.4 \\
HD\,193117 & 99.8 & 91.0 & 72.1 & 89.5 \\
HD\,194334 & 99.7 & 91.3 & 63.3 & 86.4 \\
HD\,195213 & 99.7 & 89.2 & 67.9 & 88.0 \\
\hline
\end{tabular}
\end{center}
\end{table}

The results of these simulations are listed in Table~\ref{cygnus_bias}. This table gives the percentage of systems that would be detected with the same temporal sampling as our survey if their orbital period was short, intermediate, long or covering a timescale going from 2 to more than 3000 days, respectively. This reveals that at least 72\% of the systems with periods smaller than one year should have been detected. However, between 11 and 64\% of the systems with an orbital period larger than one year could have been missed. These results show that the detection efficiency of our survey is rather good, especially for systems with periods shorter than one year, but that we may also have missed some of the long-period binary systems. These results thus emphasize that the short-period systems are easier to detect than the long-period systems with this strategy of observation.

\subsection{Binary fraction in Cygnus OB associations}
The spectroscopic analysis made on the Cygnus region includes O-type stars coming from different environments. Therefore, considering them altogether boils down to study a random sample of O-stars. It is also possible to focus on each association or even on young open clusters belonging to these associations. However, in the latter case, we would deal with small number statistics. The binary fraction will thus be discussed both over the entire region and for each OB association. 
 
\begin{figure*}[ht!]
\centering
\includegraphics[width=6cm, bb=37 11 525 395,clip]{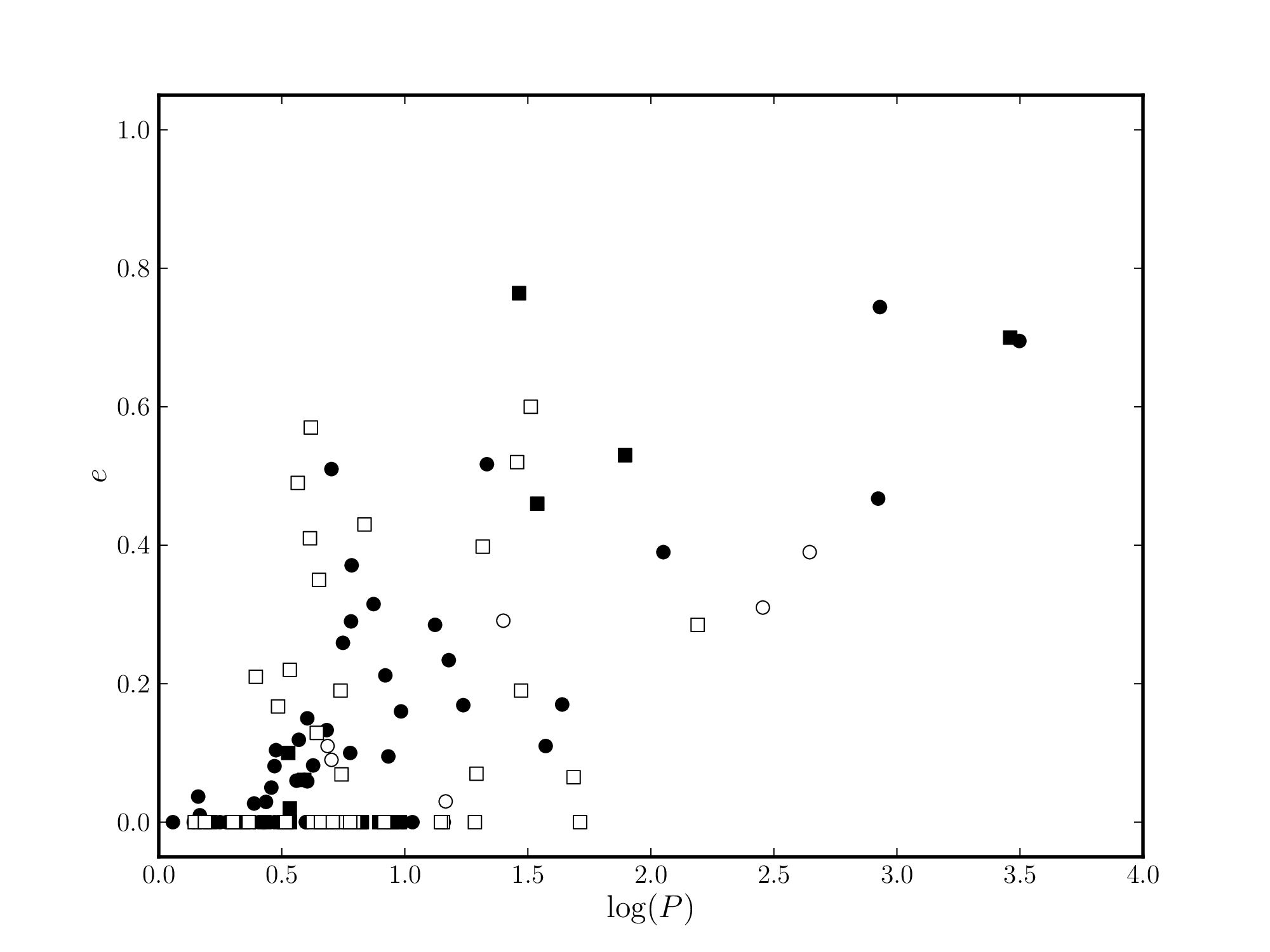}
\includegraphics[width=6cm, bb=37 11 525 395,clip]{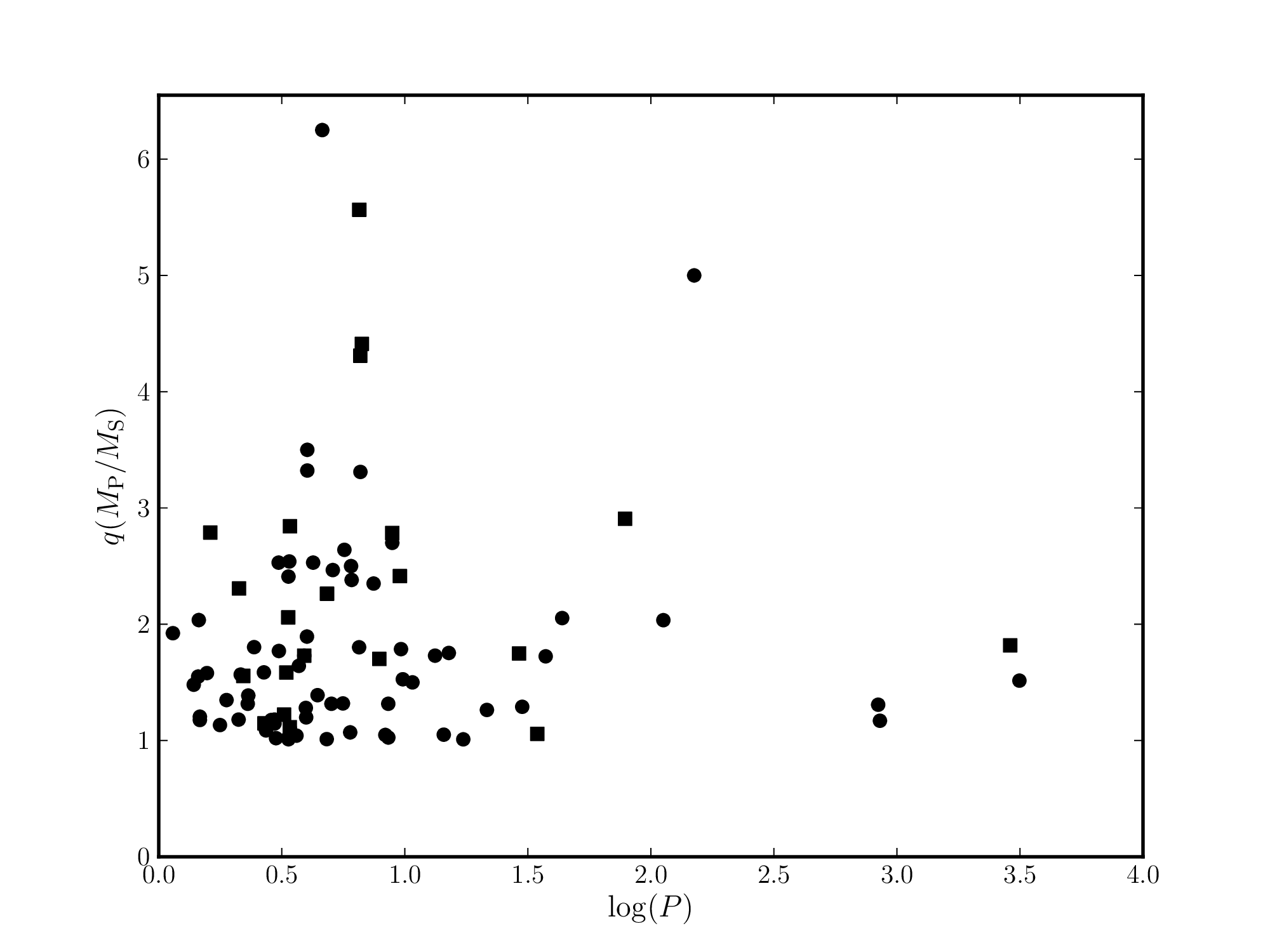}
\includegraphics[width=6cm, bb=37 11 525 395,clip]{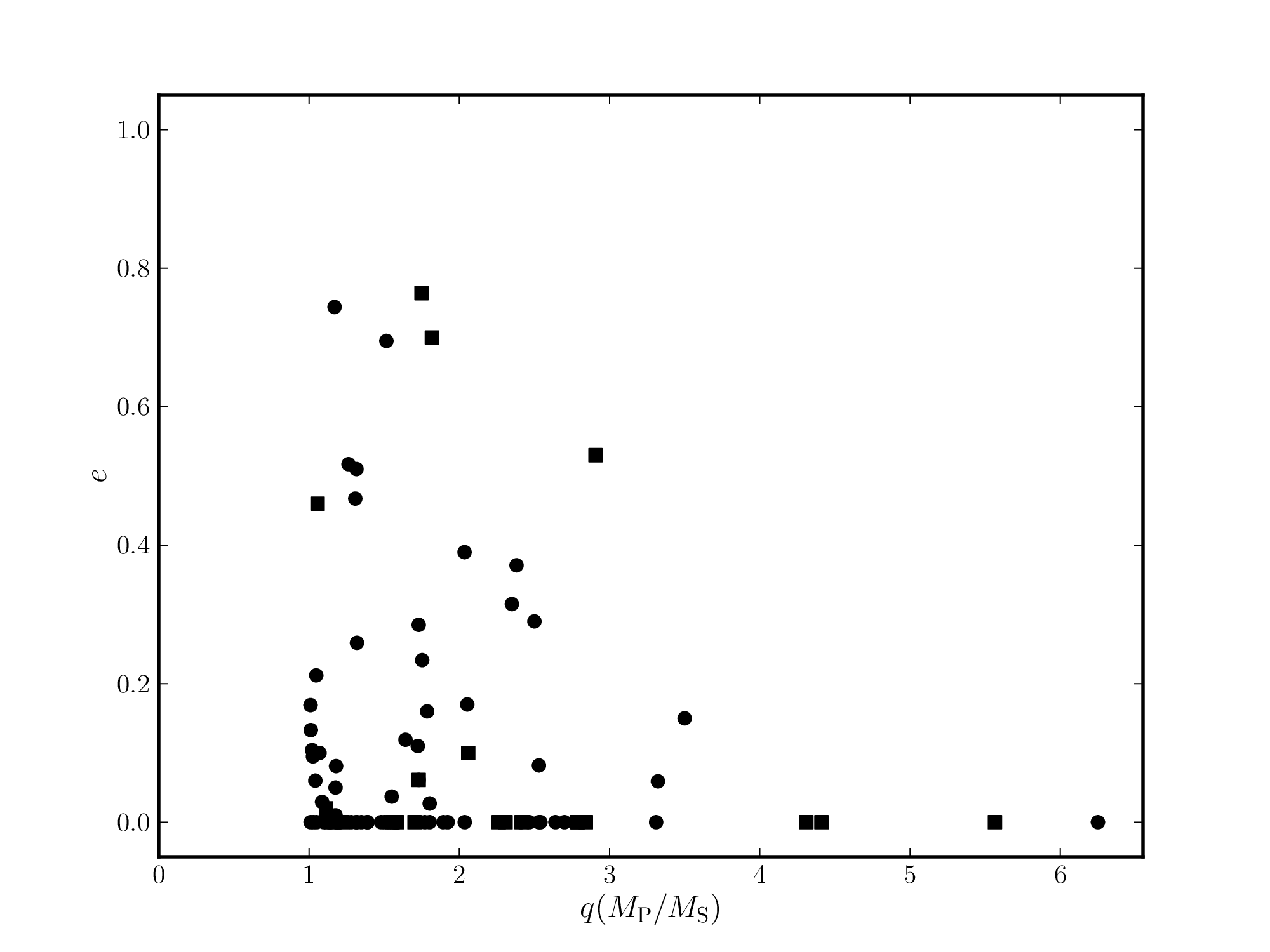}
\caption{Distributions of orbital parameters (period, eccentricity, mass ratio) of binary systems located in young open clusters or OB associations.}\label{fig:diagram}
\end{figure*}

We observed nineteen stars or multiple systems containing at least one O-type object. Among this sample, one star, HD\,229234, was classified as SB1 and three were reported as SB2 systems, HD\,193443, HD\,194649 and HD\,228989. Therefore, no doubt remains on the binary status of these objects. As a first step, we focus on each individual OB association by first including only our sample stars. In Cyg\,OB1, the minimal binary fraction is of 33\% (3 stars out of 9). In Cyg\,OB3, we find no star in a binary system out of 4, i.e., 0\% whilst in Cyg\,OB8 and Cyg\,OB9, we obtain 0\% (0 out of 3) and 33\% (1 out of 3), respectively. By putting together these numbers to achieve a more global view of our sample, we find a minimum binary fraction of about 21\% among the nineteen sample stars (4 out of 19). However, if we take into account all the O stars mentioned by \citet{hum78}, the minimum O-type star binary fractions become equal to 33\% (4 out of 12, BD\,$+36\,4063$ also detected as binary with $P_{\mathrm{orb}} = 4.8$~days, \citealt{wil09}), 33\% (3 out of 9, HD\,191201, \citealt{bur97}, HD\,190918, \citealt{hil95}, and HD\,226868, \citealt{gie82}, were detected as binaries with orbital periods of~8.3,~112.4,~5.6 days, respectively), 0\% (0 out of 4) and 14\% (1 out of 7) for the Cyg\,OB1, Cyg\,OB3, Cyg\,OB8 and Cyg\,OB9 associations, respectively. Overall, we thus obtain a minimal binary fraction of 25\% (8 stars out of 32). Of course, this value was determined by focusing only on the Cyg\,OB1, Cyg\,OB3, Cyg\,OB8 and Cyg\,OB9 associations and is therefore not representative of the entire Cygnus region. Therefore, we include in this discussion the results of the large spectroscopic/photometric survey of \citet[][and the subsequent papers]{kim12}. These authors have indeed analysed 114 stars belonging to the Cyg\,OB2 association. This sample takes into account massive objects for which the primary stars are classified between B\,3 and O\,5. They found a hard minimum binary fraction of 21\% (24 out of 114). Although our sample is clearly smaller, the minimum binary fraction appears to be similar in most of the OB associations in the Cygnus complex. Finally, putting all these results together, we reach a general minimum binary fraction of 22\% (32 out of 146) in the Cygnus region. 

\subsection{Massive binaries in a wider context}

To bring additional constraints to this work in a larger frame, we have selected the SB1 and SB2 systems (including at least one O star) detected in various spectroscopic multiplicity studies \citep{san08,san09,san11,rau04,deb06,hillwig2006,rau09,mah09} as well as in numerous papers \citep[see, e.g., "spectroscopic" references in][]{mas98} including studies devoted to the Cyg\,OB2 association \citep[][and the subsequent papers]{rau99,Let8a,cyg9bin,kim12b}. Finally, we complete this dataset by adding the orbital information concerning the O-type stars provided by the 9$^{\mathrm{th}}$ spectroscopic binary catalogue \citep{pourbaix2004}. Orbital solutions are then known for more than 140 systems (SB1\footnote{SB1 systems have not been included in the $P$ vs. $q$ and $q$ vs. $e$ plots because of the difficulty to infer reliable mass ratios in such systems.} and SB2) which are mainly located in young open clusters or OB associations. We present, in Fig.~\ref{fig:diagram}, the corresponding $P$ vs. $e$, $P$ vs. $q~(= M_{\mathrm{P}}/M_{\mathrm{S}})$ and $q$ vs. $e$ diagrams. In this figure, the filled symbols represent the SB2s, the open ones the SB1 systems, and the squares give the parameters quoted in the 9$^{\mathrm{th}}$ spectroscopic binary catalogue. In order not to affect the readibility of these diagrams and because some are missing, we do not include the error-bars on these measurements.

From the period-eccentricity diagram (left panel of Fig.~\ref{fig:diagram}), we see that shorter period systems are dominantly characterized by lower eccentricities, suggesting a lack of highly eccentric short period systems as well as almost circular long period systems. The majority of the datapoints seems to show a trend between the period and the eccentricity. However, the question of the completeness of the data has to be asked. Indeed, \citet{sanaevans2010} mentioned that at least half the known and suspected spectroscopic binaries lack a reliable orbital solution. Among the systems that lack a reliable solution, the majority has long orbital periods. The apparent linear trend between eccentricity and orbital period could be affected by two competing observational biases. Indeed, highly eccentric SB2 systems display large RV separations at periastron which are easier seen than the lower RV excursions of circular systems. However the duration over which these large separations are observed is rather short and could easily be missed. 

We see from the $P$ vs. $q$ plot (middle panel of Fig.~\ref{fig:diagram}) that the majority of detected systems are those with a period smaller than $30$~days and with a mass ratio between $1$ and $3$. This plot does not agree with a uniform distribution of the mass ratios in the range $1<q<5$ as \citet{sanaevans2010} suggested it. However, these authors also reported a decrease of the number of systems with $q>1.7$ as it can be see in Fig.~\ref{fig:diagram}. These results clearly show the limitations of spectroscopy. Indeed, it is hardly possible to detect a secondary signature for systems with large mass differences without having high-quality high signal-to-noise data. 

Finally, the $q$ vs. $e$ plot (right panel of Fig.~\ref{fig:diagram}) shows that a larger amount of systems with small eccentricities have mass ratios between $1$ and $3$ or said differently that only systems with almost similar components display a small eccentricity. However, the detectability of systems with a high mass ratio or a high eccentricity is difficult. Our sample could thus be biased. The decreasing range of high eccentricities for higher mass ratios, observed in the right panel of Fig.~\ref{fig:diagram} could thus likely be stemmed from decreasing completeness, making this conclusion less reliable. Indeed, the time required for the circularization, given by \citet{Hurley2002}, is dependent on the mass ratio ($q=(M_{\mathrm{P}}/M_{\mathrm{S}})$) squared. Therefore, when $q$ is very high, the theory predicts that the time to circularize the system will be larger, which is not the case here.

Although this sample remains small in comparison to the overall population of binaries among the O-type stars, these three diagrams provide a first approximation of the general distribution of orbital parameters. It would be interesting however to include in these diagrams binaries which would be detected by adaptive optics, speckle interferometry or interferometry to cover larger orbital parameter distributions.


\section{Conclusion}
\label{sect:conc}

We revisited the binary status of nineteen O-type stars located in different OB associations of the Cygnus region. We confirm the binarity of four objects, three SB2s: HD\,193443, HD\,194649, HD\,228989 and one SB1: HD\,229234. All these systems have short-term orbital periods, less than 10 days. We also derived for the first time the orbital parameters for the three SB2 systems. The apparent lack of intermediate and long period systems in these OB associations contrasts with the case of the NGC\,2244 young open cluster where only one longer period system HD\,46149 has been identified \citep[HD\,46573 being located in Mon\,OB2 association and not in NGC\,2244,][]{mah09}. However, this difference seems linked to the sample. Indeed, among the four stars belonging to Cygnus OB associations which were already known as binaries and thus not included in our sample, HD\,190918 is also a long period binary \citep[$P_{\mathrm{orb}} = 112.4$~days, ][]{hil95}. This result therefore does not allow us to point to different conditions for massive star formation.

\begin{table*}[htbp]
\footnotesize
\begin{center}
\caption{Summary of the multiplicity of stars and of their line profile variations}\label{tab:sum}
\begin{tabular}{lccccc}
\hline\hline
Stars	&   Sp. type    & \multicolumn{4}{c}{TVS variations} \\	    
        &               & \ion{He}{i}~4471   & \ion{He}{ii}~4542 &  \ion{He}{ii}~4686 & \ion{He}{i}~4713   \\
\hline
HD\,193443 & O\,9 $+$ O\,9.5     & Y & Y & Y & Y \\
HD\,193514 & O\,7--O\,7.5 III(f) & Y & Y & Y & N \\
HD\,193595 & O\,7V & Y & N & N & N \\
HD\,193682 & O\,5III(f)& N & N & Y & N \\
HD\,194094 & O\,8III& -- & -- & -- & -- \\
HD\,194280 & O\,9.7& Y & N & N & N \\
HD\,228841 & O\,7n  & N  & N & N & N  \\
HD\,228989 & O\,8.5V $+$ O\,9.7V & Y & Y & Y & Y \\
HD\,229234 & O\,9III $+$ $...$      & Y & Y & Y & Y \\
HD\,190864 & O\,6.5III(f) & Y & N & Y & N \\
HD\,227018 & O\,6.5V((f)) & Y & N & N & N \\
HD\,227245 & O\,7V--III((f))& -- & -- & -- & -- \\
HD\,227757 & O\,9V& -- & -- & -- & -- \\
HD\,191423 & ON\,9IIIn& N & N & N & N \\
HD\,191978 & O\,8III& Y & N & Y & N \\
HD\,193117 & O\,9III& Y & N & N & N \\
HD\,194334 & O\,7--O\,7.5 III(f)& Y & N & Y & N \\
HD\,194649 & O\,6III(f) $+$ O8V  & Y & Y & Y & Y \\
HD\,195213 & O\,7III(f) & Y & N & Y & N \\
\hline
\end{tabular}
\tablefoot{
Notes: ``Y'' means that significant TVS variations in the spectral lines are observed whilst ``N'' reports the absence of variations.
}
\end{center}
\end{table*}

The stars contained in our sample were chosen because of their brightness and because, for most of them, their binary status has not yet been established. These nineteen stars thus constitute a random sample of O-type stars. Our strategy of observations allowed us to reach a binary detection rate close to 90\% for periods up to 3165 days and to determine a minimum spectroscopic binary fraction in this sample of 21\%. By including the stars of Humphreys' catalogue (1978) which belong to the OB associations studied in the present paper but which were not analysed because their binarity was already known, we reach a minimum spectroscopic binary fraction of 25\%. When we consider all these associations separately, we obtain 33\% of O-type stars in binary systems in Cyg\,OB1, 33\% in Cyg\,OB3, 0\% in Cyg\,OB8 and 14\% in Cyg\,OB9. All these values can be completed by the minimum binary fraction of 21\% quoted by \citet{kim12} in Cyg\,OB2. Finally, when we take all these binary fractions together, we reach a binary fraction for the Cygnus region of about 22\% (32 stars out of 146).

In addition to the analysis of the RV variations, several objects show significant variations of their line profiles. These variations are mainly observed for the \ion{He}{i}~4471 and \ion{He}{ii}~4686 lines. Since not all the lines are affected, this implies that these variations are probably not due to binarity but rather result from phenomena intrinsic to stellar winds or perhaps even from non-radial pulsations. The spectral classifications of the stars derived in the present paper and the results of the TVS analysis are summarized in Table~\ref{tab:sum}.

The results presented here only focus on a small sample of O-type stars in the Cygnus complex. This region represents a large panel of O-type stars which can provide numerous clues on the formation scenarios of these objects. In paper II, we will continue the analysis of these stars by determining with a model atmosphere code their individual parameters, their age, and their chemical enrichment to better constrain the properties of these objects.

\begin{acknowledgements}
We thank the anonymous referee for useful remarks and comments. This research was supported by the PRODEX XMM/Integral contract (Belspo), the Fonds National de la Recherche Scientifique (F.N.R.S.) and the Communaut\'e fran\c caise de Belgique -- Action de recherche concert\'ee -- A.R.C. -- Acad\'emie Wallonie-Europe. We also thank the staff of San Pedro M{\`a}rtir Observatory (Mexico) and of Observatoire de Haute-Provence (France) for their technical support. The SIMBAD database has been consulted for the bibliography.
\end{acknowledgements}

\bibliography{laurent.bib}

\end{document}